\documentclass{aa} % for a paper on 1 column
\usepackage{natbib}
\usepackage{graphicx}
\usepackage{txfonts}
\usepackage{epsfig}
\usepackage{times}
\usepackage{rotating}
\usepackage{setspace}
\usepackage{lscape}
\def\aFe{[$\alpha/{\rm Fe}$]}

\def\Hb{${\rm H}{\beta}$}
\def\Mgb{{\rm Mg}$b$}
\def\Fe{$\langle {\rm Fe}\rangle$} 
\def\FeH{[{\rm Fe}$/${\rm H}]}
\def\ZH{[$Z/{\rm H}$]}
\def\MgFe{[${\rm MgFe}$]$'$}
\def\Mgd{{\rm Mg}$_2$}

\def\oiiiqc{[\ion{O}{iii}]$\lambda\lambda4959,5007$}

\def\oiiic{[\ion{O}{iii}]$\lambda5007$}
\def\niqc{[\ion{N}{i}]$\lambda\lambda5198,5200$}

\def\kms{$\rm km\;s^{-1}$}
\def\kmsmpc{$\rm km\;s^{-1}\;Mpc^{-1}$}
\def\msun{M$_\odot$}

\def\spose#1{\hbox to 0pt{#1\hss}}

\def\aj{AJ}                   % Astronomical Journal
\def\apj{ApJ}                 % Astrophysical Journal
\def\apjl{ApJ}                % Astrophysical Journal, Letters
\def\apjs{ApJS}               % Astrophysical Journal, Supplement
\def\aap{A\&A}                % Astronomy and Astrophysics
\def\aaps{A\&AS}              % Astronomy and Astrophysics, Supplement
\def\mnras{MNRAS}             % Monthly Notices of the RAS
\def\pasp{PASP}               % Publications of the ASP
\def\nat{Nature}              % Nature
\begin{document}

\title{Fossil group origins}

\subtitle{IX. Probing the formation of fossil galaxy groups with
  stellar population gradients of their central galaxies\thanks{Based
    on observations made with the Gran Telescopio Canarias (GTC),
    installed in the Spanish Observatorio del Roque de los Muchachos
    of the Instituto de Astrof\'\i sica de Canarias
    (IAC).}$^,$\thanks{Tables \ref{tab:ellipse}, \ref{tab:kinematics},
    and \ref{tab:indices} are only available in electronic form at the
    CDS via anonymous ftp to cdsarc.u-strasbg.fr (130.79.128.5) or via
    http://cdsweb.u-strasbg.fr/cgi-bin/qcat?J/A+A/.}}

\authorrunning{E. M. Corsini et al.}

\titlerunning{Stellar populations of fossil groups in central galaxies}

\author{E. M. Corsini\inst{1,2}, 
L. Morelli\inst{1,2,3},
S. Zarattini\inst{4,5,6},
J. A. L. Aguerri\inst{7,8},
L. Costantin\inst{1}, 
E. D'Onghia\inst{9,10},
M. Girardi \inst{4,5}, \\
A. Kundert\inst{9}, 
J. M\'endez-Abreu\inst{7,8}, and
J. Thomas\inst{11}} 

\institute{Dipartimento di  Fisica e Astronomia ``G. Galilei'', Universit\`a di Padova,
              vicolo dell'Osservatorio 3, I-35122 Padova, Italy\\
              \email{enricomaria.corsini@unipd.it}
           \and
           INAF--Osservatorio Astronomico di Padova, vicolo dell'Osservatorio~2, 
           I-35122 Padova, Italy
           \and 
           Instituto de Astronom\'\i a y Ciencias Planetarias,
           Universidad de Atacama, Copayapu 485, Copiap\'o, Chile
           \and
           Dipartimento di Fisica--Sezione Astronomia, Universit\`a di Trieste, 
           via Tiepolo 11, I-34143 Trieste, Italy
           \and 
           INAF--Osservatorio Astronomico di Trieste, via Tiepolo 11, 
           I-34143 Trieste, Italy
           \and        
           Institut de Recherche sur les Lois Fondamentales de
           l'Univers, Commissariat \`a l'\'Energie Atomique,
           Universit\'e Paris-Saclay, F-91191 Gif-sur-Yvette, France
           \and
           Instituto de Astrof\'\i sica de Canarias, calle V\'\i a L\'actea s/n, 
           E-38205 La Laguna, Tenerife, Spain
           \and 
           Departamento de Astrof\'\i sica, Universidad de La Laguna, 
           Avenida Astrof\'\i sico Francisco S\'anchez s/n, E-38206 La Laguna, 
           Tenerife, Spain
           \and 
           Astronomy Department, University of Wisconsin, 475 Charter St., 
           Madison, WI 53706, USA
           \and 
           Center for Computational Astrophysics, Flatiron Institute,
           162 Fifth Avenue, New York, NY 10010, USA
           \and 
           Max-Planck-Institut f\"ur extraterrestrische Physik, Giessenbachstra\ss e, 
           D-85748 Garching, Germany}

\date{\today}

\abstract{Fossil groups (FGs) are galaxy aggregates with an extended
  and luminous X-ray halo, which are dominated by a very massive
  early-type galaxy and lack of $L^\ast$ objects. FGs are indeed
  characterized by a large magnitude gap between their central and
  surrounding galaxies. This is explained by either speculating that
  FGs are failed groups that formed without bright satellite galaxies
  and did not suffer any major merger, or by suggesting that FGs are very
  old systems that had enough time to exhaust their bright satellite
  galaxies through multiple major mergers.}
{Since major mergers leave signatures in the stellar populations of
  the resulting galaxy, we study the stellar population parameters of
  the brightest central galaxies (BCGs) of FGs as a benchmark against
  which the formation and evolution scenarios of FGs can be compared.}
{We present long-slit spectroscopic observations along the major,
  minor, and diagonal axes of NGC~6482 and NGC~7556, which are the
  BCGs of two nearby FGs. The measurements include spatially resolved
  stellar kinematics and radial profiles of line-strength 
  indices, which we converted into stellar population parameters using
  single stellar-population models.}
{NGC~6482 and NGC~7556 are very massive ($M_\ast\simeq10^{11.5}$
  \msun) and large ($D_{25}\simeq50$ kpc) galaxies. They host a
  centrally concentrated stellar population, which is significantly
  younger and more metal rich than the rest of the galaxy. The age
  gradients of both galaxies are somewhat larger than those of the
  other FG BCGs studied so far, whereas their metallicity gradients
  are similarly negative and shallow. Moreover, they have negligible
  gradients of $\alpha$-element abundance ratio.}
{The measured metallicity gradients are less steep than those
  predicted for massive galaxies that formed monolithically and
  evolved without experiencing any major merger. We conclude that the
  observed FGs formed through major mergers rather than being failed
  groups that lacked bright satellite galaxies from the beginning.}


\maketitle

\section{Introduction}

Fossil groups (FGs) were first proposed by \citet{Ponman1994}, when
they found that the apparently isolated galaxy RX~J1340.6$+$4018 was
surrounded by an X-ray halo typical of a galaxy group. They suggested
that this was a fossil relic of an old galaxy group that had enough
time to merge all its bright satellite galaxies into the central
one. In this scenario, FGs formed at high redshift ($z > 1$), with few
subsequent accretions \citep{DOnghia2005}. If these systems really
exist today, they can offer a unique opportunity to directly study the
formation processes that occurred several gigayears ago, thus shedding
light on the assembly of massive halos in the early universe.

%%%%%%%%%%%%%%%%%%%%%%%%%%%%%%%%%%%%%%%%%%%%%%%%%%%%%%%%%%%%%%%%%%%% 
% TABLE: SAMPLE
\renewcommand{\tabcolsep}{3pt}
\begin{table*}[t!]  
\caption{Properties of the sample galaxies. \label{tab:sample}}
\begin{center}
\begin{small}
\begin{tabular}{llccccccc}    
\hline
\noalign{\smallskip}
\multicolumn{1}{c}{Galaxy} & 
\multicolumn{1}{c}{Type} & 
\multicolumn{1}{c}{$z$} &
\multicolumn{1}{c}{$D_A$} &
\multicolumn{1}{c}{scale} &
\multicolumn{1}{c}{$D_{25}$} &
\multicolumn{1}{c}{$(m-M)_L$} &
\multicolumn{1}{c}{$M_K$} & 
\multicolumn{1}{c}{$\log{M_\ast}$} \\
\multicolumn{1}{c}{[name]} & 
\multicolumn{1}{c}{} & 
\multicolumn{1}{c}{} &
\multicolumn{1}{c}{[Mpc]} &
\multicolumn{1}{c}{[pc arcsec$^{-1}$]} &
\multicolumn{1}{c}{[kpc]} &
\multicolumn{1}{c}{[mag]} &
\multicolumn{1}{c}{[mag]} &
\multicolumn{1}{c}{[$\log$\msun]}\\
\multicolumn{1}{c}{(1)} & 
\multicolumn{1}{c}{(2)} & 
\multicolumn{1}{c}{(3)} &
\multicolumn{1}{c}{(4)} &
\multicolumn{1}{c}{(5)} &
\multicolumn{1}{c}{(6)} &
\multicolumn{1}{c}{(7)} &
\multicolumn{1}{c}{(8)} &
\multicolumn{1}{c}{(9)} \\
\noalign{\smallskip}
\hline
\noalign{\smallskip}
NGC~6482 & E:     & 0.013 & 54.3 & 263 & 31 & 33.73 & $-25.36$ & 11.41\\
NGC~7556 & S0$^-$ & 0.025 & 99.2 & 481 & 72 & 34.98 & $-25.69$ & 11.54\\
\noalign{\smallskip}
\hline  
\end{tabular}
\end{small}
\end{center}
\tablefoot{(1) Galaxy name. (2) Morphological type from
  \citet[][RC3]{RC3}.  (3) Redshift corrected to the cosmic microwave
  background reference frame from the NASA/IPAC Extragalactic Database
  (NED). (4) Angular distance from NED. (5) Conversion factor from
  arcsec to parsec.  (6) Major-axis diameter of the isophote at 
  surface brightness level $\mu_B=25$ mag arcsec$^{-2}$ from RC3.
  (7) Distance modulus from the luminosity distance in NED. (8)
  Absolute magnitude in the $K_s$ band from the apparent magnitude in
  the Extend Source Catalog of the Two Micron All Sky Survey
  \citep{Jarrett2000} and after applying the Galactic absorption
  correction \citep{Schlafly2011} and $K$ correction
  \citep{Chilingarian2010} available in NED. (9) Stellar mass from
  $M_K$. We derived the mass-to-light ratio following \citet{Bell2003}
  and adopting $B-R$ \citep{Lieder2013} for NGC~6482 and $g-r$
  \citep{Alam2015} for NGC~7556 by taking into account Galactic
  absorption and $K$ correction.}
\end{table*}
%%%%%%%%%%%%%%%%%%%%%%%%%%%%%%%%%%%%%%%%%%%%%%%%%%%%%%%%%%%%%%%%%%%% 

FGs are defined, from an observational point of view, as groups or
clusters of galaxies dominated by a large early-type galaxy (ETG), in
which the second-ranked member within half the virial radius is at
least two magnitudes fainter in the $r$-band \citep[$\Delta m_{12} >
  2$, ][]{Jones2003}. Moreover, an extended X-ray halo of at least
$L_X > 10^{42}$ $h_{50}$ erg s$^{-1}$ must be detected in order to
avoid the classification of large isolated galaxies as FGs. The lack
of $L^*$ galaxies is thought to be the result of the cannibalism of
the central galaxy. In fact, dynamical friction is the process
responsible for the merging of galaxies and it is more effective when
massive galaxies are involved. Furthermore, the relaxed dynamical
status of RX~J1340.6$+$4018 was considered an indicator of the old age
of the system. The observational definition was slightly modified by
\citet{Dariush2010}, who proposed to use the gap in magnitude between
the first and the fourth brightest galaxies ($\Delta m_{14}$) as a
reference. In particular, they suggested that $\Delta m_{14} > 2.5$
within half the virial radius was a more stable identification
criterion than the $\Delta m_{12} > 2$ criterion proposed by
\citet{Jones2003}.

Several samples of FGs selected using the $\Delta m_{12}$ or $\Delta
m_{14}$ criteria have been presented in the last decade thanks to the
availability of new surveys \citep{Khosroshahi2007, Santos2007,
  Voevodkin2010, Proctor2011, Harrison2012}. As the number of FGs
increased, their observational properties became clearer. In
particular, many studies focused on the properties of the hot
intra-cluster component \citep{Khosroshahi2007, Khosroshahi2014,
  Proctor2011, Harrison2012, Girardi2014, Kundert2015} or on the
properties of the satellite galaxy population \citep{Khosroshahi2006,
  Khosroshahi2014, MendesDeOliveira2006, Aguerri2011, Aguerri2017,
  Adami2012, Lieder2013, Zarattini2014, Zarattini2015, Zarattini2016}.

At the same time, numerical simulations show that the magnitude gap
alone is not a good indicator of the dynamical stage of a group or
cluster \citep{DOnghia2004, DOnghia2005}. Von Benda-Beckmann et
al. (2008) suggested that the fossil status is just a stage in the
evolution of a system. This result has been recently confirmed by
\citet{Kundert2017}, who showed how the fossil status changes every
$2-3$ Gyr. Moreover, there are hints that FGs suffered the last major
mergers more recently than non-FGs \citep{Diaz-Gimenez2008,
  Kundert2017}. On the other hand, \citet{Gozaliasl2014} pointed out
that there is a difference in the evolution of the faint end of the
luminosity function (LF) in FGs and non-FGs. In particular, the faint
end of the LF of FGs suffered no evolution after $z \sim 1$, whereas
non-FGs went through an intense evolution, as also confirmed by
\citet{Kundert2017}. They found that the main difference between FGs
and non-FGs is the halo accretion history in the last few gigayears.

The formation and evolution of the brightest cluster galaxies (BCGs)
has also been a main topic in the study of FGs. Indeed, BCGs are a
fundamental component of a galaxy group or cluster. They are often
located close to the X-ray center of the cluster \citep{Jones1984,
  Lin2004, Lauer2014}, and they can be as luminous as $10 L^*$, where
$L^*$ is the characteristic luminosity of the cluster LF
\citep[e.g.,][]{Schombert1986}. Their growth is expected to be driven
by wet mergers if they formed at high redshift, whereas their
formation at lower redshifts could be driven by dry mergers. This
should leave an imprint on the stellar populations of BCGs, in
particular in those residing in old and relaxed FGs. Indeed,
\citet{Khosroshahi2006} found differences in the isophotal shapes of
BCGs in FGs compatible with a formation at high redshift and via wet
mergers. They found that the isophotes were more disky in FGs
than in non-FGs. Although \citet{MendezAbreu2012} did not confirm such
a difference, they demonstrated using the scaling relations of the
BCGs that they assembled part of their mass at high redshift and via
wet mergers. At the same time, they suggested that the majority of the
mass was assembled at lower redshift via dry mergers.  As a
consequence, the hints about the relative importance of the different
kinds of mergers in shaping FG BCGs appear to be quite controversial.
However, the BCGs in FGs are amongst the most massive and luminous
galaxies observed in the universe \citep{MendezAbreu2012, Kundert2017}
and this means that the merging process in FGs had to be particularly
efficient.

The stellar populations of the BCGs could help in shedding light on
the possible differences in the formation processes of FGs and
non-FGs. \citet{LaBarbera2009} found a striking similarity between the
stellar population properties of FG BCGs and bright field galaxies
indicating that they had similar star formation histories. This
observational result is in agreement with subsequent findings obtained
by \citet{Cui2011} from cosmological simulations and galaxy formation
models. They showed that FG BCGs have similar properties to those in
non-FGs in terms of age, metallicity, color, concentration, and total
mass-to-light ratio.
Results along these lines were also reported for compact
\citep{Proctor2004, MendesDeOliveira2005} and large-gap groups
\citep{Trevisan2017}, which are somehow related to FGs.
According to \citet{Proctor2004} and \citet{MendesDeOliveira2005}, the
majority of galaxies in compact groups, which are possible progenitors
of FGs, are older than those in the field but have similar
metallicities and $\alpha$-element abundance ratios. However, the
connection between compact groups and FGs is not straightforward
because they usually follow different evolutionary tracks, as recently
investigated by \citet{Farhang2017} with semi-analytic models of
galaxy formation.
\citet{Trevisan2017} compared the stellar populations of the first and
second brightest group galaxies as a function of their magnitude gap
using a complete sample of groups. The magnitude gap does not
correlate with the age, metallicity, $\alpha$-element abundance ratio,
and star formation history. Although many large-gap groups with
$\Delta m_{12}>2$ are expected to be FGs, their true link has not yet
been addressed.

All the previous studies investigated the integrated stellar
population properties of BCGs in FGs, except for
\citet{Eigenthaler2013} and \citet{Proctor2014}.
\citet{Eigenthaler2013} analyzed spatially resolved stellar
populations in a sample of six BCGs in FGs: they concluded that FGs
formed via the merging of the $L^*$ galaxies with the central one,
excluding the monolithic collapse in which the magnitude gap would not
be produced by evolutionary effect but would have been in place {\it a
  priori}. \citet{Proctor2014} performed a similar analysis on a
sample of two FGs finding that, despite remarkable similarities in
their morphology, photometric properties, and kinematics, the stellar
populations of the two galaxies were clearly different. One shows a
strong gradient all the way to the center, with signs of a burst of
stellar formation located in the center of the galaxy and superimposed
onto an old and extended population. On the contrary, the second
galaxy of their sample showed a flat core-like structure in the
metallicity gradient, but no age gradient.

This paper is part of the Fossil Group Origins (FOGO) project
\citep{Aguerri2011}, whose aim was to study a large sample of FG
candidates, spanning wide ranges in mass and redshift by using
multi-wavelength observations. To date, the collaboration has
published results on the properties of the BCGs
\citep{MendezAbreu2012, Zarattini2014}, on their dark matter halos
\citep{Girardi2014, Kundert2015}, on the galaxy population
\citep{Zarattini2015, Zarattini2016, Aguerri2017}, and on the
comparison between observations and current cosmological simulations
\citep{Kundert2017}. Here, we focus our attention on the stellar
populations of two BCGs in FGs, namely NGC~6482 and NGC~7556.  They
are interesting because there are only a few nearby FG BCGs that are
sufficiently bright at radii larger than the effective radius to
obtain in a reasonable amount of time the high signal-to-noise ratio
spectra needed for measuring their stellar population gradients.  We
aim at increasing the number of FG BCGs for which age, metallicity,
and $\alpha$-element abundance ratio are known at the same level of
detail as those of BCGs in non-FGs \citep[e.g.,][]{Mehlert2003,
  Brough2007, Loubser2012}. A reliable comparison between the stellar
populations of FG and non-FG BCGs is highly desirable, but not yet
possible given the current small number statistics of FG BCGs.

The paper is organized as follows. We present the target galaxies in
Sect.~\ref{sec:sample}, and we describe the analysis of their
photometric and spectroscopic data in Sects.~\ref{sec:photometry} and
\ref{sec:spectroscopy}, respectively. We derive the stellar population
properties in Sect.~\ref{sec:populations}. We discuss the results in
Sect.~\ref{sec:discussion} and give our conclusions in
Sect.~\ref{sec:conclusions}. We adopt $H_0 = 70$ \kmsmpc ,
$\Omega_{\rm m} = 0.3$, and $\Omega_\Lambda = 0.7$ as cosmological
parameters all throughout the paper.

\section{Sample selection}
\label{sec:sample}

The selected galaxies, whose main properties are listed in
Table~\ref{tab:sample}, are the BCGs of two nearby FGs with X-ray
observations. 

NGC~6482 is the BCG of MCXC~J1751.7+2304, which is the closest known
FG at $z=0.013$. It is a non-cool core system with an X-ray luminosity
$L_X = 1.0 \times 10^{42}$ erg s$^{-1}$, a virial mass $M_{200}=4.0
\times 10^{12}$ M$_\odot$, and a virial radius $R_{200} = 310$ kpc
\citep{Khosroshahi2004}.  The magnitude gap between the two brightest
member galaxies is $\Delta m_{12} = 2.19\pm0.04$ mag
\citep{Lieder2013}.

NGC~7556 is the dominant galaxy of RXC~J2315.7$-$0222 at a redshift of
$z=0.025$. This system is characterized by a cool core and it has an
X-ray luminosity $L_X = 2.1 \times 10^{43}$ erg s$^{-1}$, a mass
$M_{500} = 5.4 \times 10^{13}$ M$_\odot$, and a radius $R_{500} = 569$
kpc \citep{Democles2010}. We derived its virial mass $M_{200} = 7.4
\times 10^{13}$ M$_\odot$ and virial radius $R_{200} = 859$
kpc. RXC~J2315.7$-$0222 is actually a {\em bona fide} FG, since its
magnitude gap is $\Delta m_{12} = 1.88\pm0.03$ mag
\citep{Democles2010}.  Nevertheless, we decided to analyze it due to
the paucity of nearby FGs.

\section{Broad-band photometry}
\label{sec:photometry}

We need the effective radius of NGC~6482 and NGC~7556 to derive the
central values and radial gradients of their stellar population
parameters. 

For NGC~6482 we adopted the structural parameters obtained by
\citet{Lieder2013}.  They performed a deep $R$-band imaging of
NGC~6482 with the Suprime camera mounted at the 8.2-m Subaru telescope
and modeled the surface brightness distribution of the galaxy with a
S\'ersic law (see Eq.~\ref{eq:sersic}). The parameters are reported in
Table~\ref{tab:parameters}.  

For NGC~7556 we retrieved the flux-calibrated $r$-band image available
in the Data Release 12 of the Sloan Digital Sky Survey
\citep[SDSS-DR12,][]{Alam2015} and analyzed it to obtain the
structural parameters of the galaxy.

%%%%%%%%%%%%%%%%%%%%%%%%%%%%%%%%%%%%%%%%%%%%%%%%%%%%%%%%%%%%%%%%%%%% 
% TABLE: STRUCTURAL PARAMETERS
\renewcommand{\tabcolsep}{3pt}
\begin{table}  
\caption{Structural parameters of the sample galaxies.
\label{tab:parameters}}
\begin{center}
\begin{small}
\begin{tabular}{lll}    
\hline 
\noalign{\smallskip}   
\multicolumn{1}{c}{Parameter} &  
\multicolumn{1}{c}{NGC~6482} & 
\multicolumn{1}{c}{NGC~7556} \\   
\multicolumn{1}{c}{(1)} &   
\multicolumn{1}{c}{(2)} & 
\multicolumn{1}{c}{(3)} \\   
\noalign{\smallskip}   
\hline
\noalign{\smallskip}       
$I_{\rm e}$ [mag arcsec$^{-2}$] & $19.88 \pm 0.02$ & $20.92 \pm 0.03$\\ 
$r_{\rm e}$ [arcsec]           & $16.7  \pm 0.1$  & $9.6 \pm 0.2$ \\
$n$                          & $2.82\pm0.03$    & $2.74 \pm 0.03$\\
${\it PA}_{\rm sph}$ [$^\circ$]  & \ldots           & $112.5 \pm 0.5$ \\
$q_{\rm sph}$                   & \ldots           & $0.756 \pm 0.003$\\ 
$I_{\rm 0}$ [mag arcsec$^{-2}$] & \ldots           & $21.25 \pm 0.01$ \\
$h$ [arcsec]                 & \ldots           & $28.5 \pm 0.3$ \\
${\it PA}_{\rm env}$ [$^\circ$]  & \ldots           & $117.0 \pm 0.2$ \\
$q_{\rm env}$                  & \ldots            & $0.639 \pm 0.002$\\
\noalign{\smallskip}       
\hline
\end{tabular} 
\end{small}
\end{center}     
 \tablefoot{(2) From \citet{Lieder2013}. We calculated the effective
  radius along the galaxy major axis from the circularized effective
  radius ($r_{\rm e}^{\rm c} = 14.20\pm0.09$ arcsec) and effective
  ellipticity ($\epsilon(r_{\rm e}^{\rm c})=0.28$). (3) From this
  paper.}
\end{table}    
%%%%%%%%%%%%%%%%%%%%%%%%%%%%%%%%%%%%%%%%%%%%%%%%%%%%%%%%%%%%%%%%%%%%  

\subsection{Image reduction}

We measured the sky level to be subtracted from the SDSS $r$-band
image of NGC~7556 following the procedure proposed by
\citet{Pohlen2006}, as done in \citet{Morelli2016}. We masked the
stars, galaxies, and spurious sources in the neighborhood of the
  galaxy and measured its surface brightness radial profile with the
ELLIPSE task in IRAF.{\footnote{The Image Reduction and Analysis
    Facility is distributed by the National Optical Astronomy
    Observatory (NOAO), which is operated by the AURA, Inc., under a
    cooperative agreement with the National Science Foundation.}
  First, we fitted the galaxy isophotes with ellipses having the
  center, ellipticity, and position angle free to vary. Then, we
  repeated the isophotal fit, fixing the center we previously obtained
  for the inner ellipses and the ellipticity and position angle of the
  outer ones. We calculated the sky level by averaging the surface
  brightness measured at large radii where there is no light
  contribution from the galaxy. We used the IRAF task IMEXAMINE to
  measure the standard deviation of the background in the
  sky-subtracted images and to fit the stars of the field of view with
  a circular Moffat profile \citep{Moffat1969}, which we adopted to
  model the point spread function (PSF).
Finally, we trimmed the sky-subtracted image to reduce the computing
time required to perform a reliable photometric decomposition and we
ran ELLIPSE on the trimmed images to derive the radial profiles of
surface brightness, ellipticity, and position angle to be used for the
photometric decomposition. They are given in
Table~\ref{tab:ellipse}.

%%%%%%%%%%%%%%%%%%%%%%%%%%%%%%%%%%%%%%%%%%%%%%%%%%%%%%%%%%%%%%%%%%%% 
% FIGURE: PHOTOMETRIC DECOMPOSITION NGC 7556 
\begin{figure*}[t!]
\centering
\includegraphics[angle=270, width=0.8\textwidth]{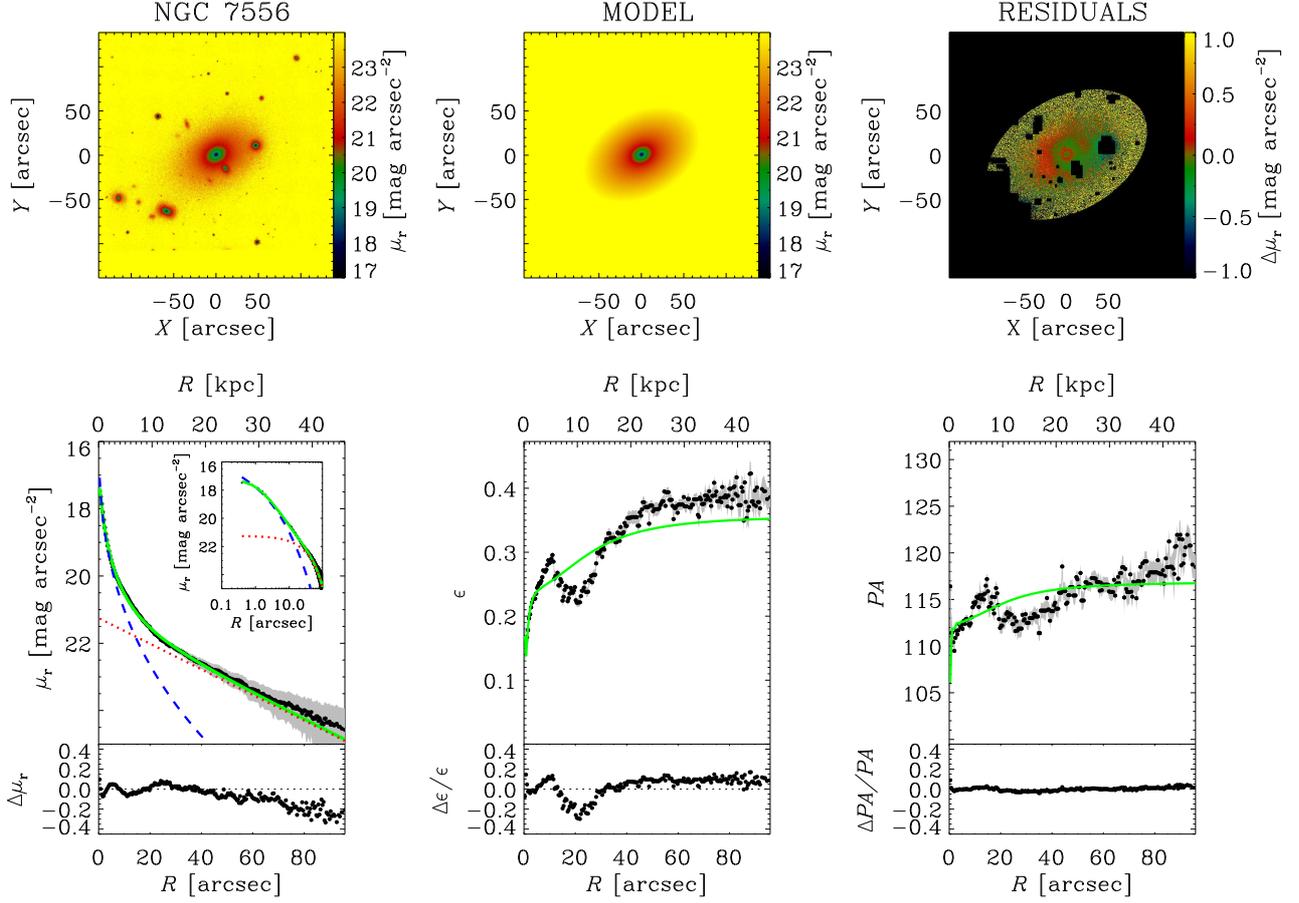}
\vspace{1cm}
\caption{Two-dimensional photometric decomposition of the $r$-band
  image of the NGC~7556. The upper panels (from left to right) show
  the map of the observed, modeled, and residual (observed$-$modeled)
  surface brightness distributions. The field of view is oriented with
  North up and East left. The black areas in the residual image
  correspond to pixels excluded from the fit. The lower panels (from
  left to right) show the ellipse-averaged radial profile of surface
  brightness, position angle, and ellipticity measured in the observed
  (black dots with gray error bars) and seeing-convolved modeled image
  (green solid line) and their corresponding difference. The intrinsic
  surface-brightness radial profiles of the best-fitting spheroid
  (blue dashed line) and halo (red dotted line) components are also
  shown in both linear and logarithmic scale for the distance to the
  center of the galaxy.}
\label{fig:decomposition}
\end{figure*}
%%%%%%%%%%%%%%%%%%%%%%%%%%%%%%%%%%%%%%%%%%%%%%%%%%%%%%%%%%%%%%%%%%%% 

\subsection{Photometric decomposition}

We performed the photometric decomposition of the sky-subtracted image
of NGC~7556 by using the Galaxy Surface Photometry 2-Dimensional
Decomposition algorithm \citep[GASP2D,][]{MendezAbreu2008,
  MendezAbreu2014}.

GASP2D assumes that the observed surface brightness in each image
pixel is expressed as the sum of analytical functions describing the
light contribution of the structural components of the
galaxy. Although classified as an early S0 galaxy \citep{RC3},
NGC~7556 shows a bright spheroidal component surrounded by a faint
extended envelope rather than a disk. This envelope contributes a
light excess at large radii over the de Vaucouleurs profile
\citep{deVaucouleurs1948} fitted to the surface brightness in the
inner parts of the galaxy. Such an excess is typical for central
dominant galaxies (cDs) in clusters \citep{Schombert1987}. Therefore,
we modeled the surface brightness of the spheroidal body of the galaxy
with a S\'ersic law \citep{Sersic1968} and the surface brightness of
the stellar envelope with an exponential law \citep{Freeman1970}.
Adopting a single S\'ersic law did not provide a reliable fit of the
galaxy surface brightness at all radii \citep[see also][and
  references therein]{Ascaso2011}. 
For the spheroidal component, the surface brightness profile is
\begin{equation}
I_{\rm sph}(r) = I_{\rm e} 10^{-b_n[(r/r_{\rm e})^{1/n} -1 ]} ,
\label{eq:sersic}
\end{equation}
where $r_{\rm e}$, $I_{\rm e}$, and $n$ are the effective radius,
effective surface brightness, and shape parameter of the surface
brightness profile, respectively, while $b_n = 0.868n - 0.142$
\citep{Caon1993} is a normalization coefficient.  For the luminous
envelope, the surface brightness profile is
\begin{equation}
I_{\rm env}(r) = I_{\rm 0} e^{-r/h} ,
\end{equation}
where $I_{\rm 0}$ and $h$ are the central surface brightness and
scalelength of the surface brightness profile, respectively. 
We assumed the isophotes of both the spheroid and envelope to be
elliptical, centered onto the galaxy center, and with constant
position angle ${\it PA}_{\rm sph}$ and ${\it PA}_{\rm env}$, and constant
axial ratio $q_{\rm sph}$ and $q_{\rm env}$, respectively. We did not
consider other components such as rings.

We obtained the best-fitting values of the structural parameters of
the spheroid and envelope by weighting the surface brightness of the
image pixels according to the variance of the total observed photon
counts due to the contribution of both galaxy and sky. Moreover,
GASP2D accounts for photon noise, charge-coupled device (CCD) gain and
read-out noise, and image PSF. We show the photometric decomposition
of NGC~7556 in Fig.~\ref{fig:decomposition} and give its structural
parameters in Table~\ref{tab:parameters}. The surface brightness
contribution of the spheroidal component is equal to the contribution
from the surrounding luminous envelope at $r_{\rm se} = 14.2$ arcsec.

We derived the errors on the structural parameters from
\citet{MendezAbreu2017}. They built a large set of images of mock
galaxies of different morphological types and analyzed them as if they
were real using GASP2D. We considered their results for the magnitude
bin $12<m_r<13$ mag since NGC~7556 has $m_r = 12.32$ \citep{Alam2015}.
For $I_{\rm e}$, $r_{\rm e}$, $n$, $I_{0}$, and $h$ we adopted the
mean and standard deviation of the relative errors of the mock
galaxies as the systematic and statistical errors of the observed
galaxies, respectively. For $q_{\rm sph}$, $q_{\rm env}$, and ${\it
  PA}_{\rm sph}$, ${\it PA}_{\rm env}$ we adopted the mean and
standard deviation of the absolute errors of the mock galaxies as the
systematic and statistical errors $\sigma_{\rm syst}$ and $\sigma_{\rm
  stat}$ of the observed galaxies, respectively. We computed the
errors as $\sigma^2$ = ${\sigma_{\rm stat}^2 + \sigma_{\rm syst}^2}$,
with the systematic errors negligible compared to the statistical
ones.

\section{Long-slit spectroscopy}
\label{sec:spectroscopy}

We measured the stellar and ionized-gas kinematics and line-strength
indices of NGC~6482 and NGC~7556 from the long-slit spectra we
obtained at the 10.4-m Gran Telescopio Canarias (GTC) telescope in La
Palma (Spain).

\subsection{Observations and spectra reduction}
\label{sec:observations}

We carried out the spectroscopic observations at GTC in service mode
between April and August 2013.
The Optical System for Imaging and low-intermediate-Resolution
Integrated Spectroscopy imager and spectrograph
\citep[OSIRIS,][]{Cabrera-Lavers2014} was used in combination with the
R2500V grism and the 1.0 arcsec $\times$ 7.4 arcmin slit. The detector
was a mosaic of two Marconi CCD42-82 chips with a gain and readout
noise of 1.18 $e^-$~ADU$^{-1}$ and 3.5 $e^-$ (rms), respectively. The
CCDs have $2048\times2048$ pixels of $15\times15$ $\rm \mu m^2$ and
are separated by a gap of 37 pixels giving a field of view of $7.0
\times 7.0$ arcmin$^2$ with a spatial scale of 0.127
arcsec~pixel$^{-1}$.
We adopted an on-chip binning of $2\times2$ pixels to cover the
wavelength range from about 4500 to about 6000 \AA\ with a reciprocal
dispersion of 1.60 \AA~pixel$^{-1}$ and a spatial scale of 0.254
arcsec pixel$^{-1}$. We derived the instrumental resolution by
measuring the full width at half maximum (FWHM) of 12 unblended
emission lines of a comparison arc-lamp spectrum after wavelength
calibration. The mean FWHM of the arc-lamp lines from different
observing nights was $2.62\pm0.01$ \AA , corresponding to $\sigma_{\rm
  inst} = 64$ \kms\ at 5185 \AA .

We obtained the spectra along the major, minor, and a diagonal axis of
both galaxies by centering the slit on their nucleus. We also observed
several spectrophotometric standard stars to calibrate the flux of the
galaxy spectra before measuring the line-strength indices. We took
spectra of the comparison arc lamp during each observing night to
perform the wavelength calibration. The seeing FWHM during the
observations was computed by fitting a circular Gaussian to the guide
star. It ranged between 0.5 and 1.2 arcsec with a median value of 1.0
arcsec. We give the integration times of the single exposures, total
integration times for each observed axis, and slit position angle of
the galaxy spectra in Table~\ref{tab:spectroscopy}.

%%%%%%%%%%%%%%%%%%%%%%%%%%%%%%%%%%%%%%%%%%%%%%%%%%%%%%%%%%%%%%%%%%%% 
% TABLE: LOG SPECTROSCOPY 
\renewcommand{\tabcolsep}{3pt}
\begin{table}  
\caption{Log of the spectroscopic observations of the sample galaxies.
\label{tab:spectroscopy}}
\begin{center}
\begin{small}
\begin{tabular}{lrccc}    
\hline 
\noalign{\smallskip}   
\multicolumn{1}{c}{Galaxy} &  
\multicolumn{1}{c}{P.A.} & \multicolumn{1}{c}{Position} &   
\multicolumn{1}{c}{Single Exp. T.} & \multicolumn{1}{c}{Total Exp. T.} \\   
\multicolumn{1}{c}{} &  
\multicolumn{1}{c}{[$^\circ$]} & \multicolumn{1}{c}{} &  
\multicolumn{1}{c}{[s]} & \multicolumn{1}{c}{(h)} \\   
\multicolumn{1}{c}{(1)} &   
\multicolumn{1}{c}{(2)} & \multicolumn{1}{c}{(3)} &   
\multicolumn{1}{c}{(4)} & \multicolumn{1}{c}{(5)} \\ 
\noalign{\smallskip}   
\hline
\noalign{\smallskip}       
NGC~6482      & $+65$ & MJ & $600+2\times1500+2\times2700$ & 2.5 \\ 
              & $+20$ & DG & $900+2\times2700$             & 1.8 \\ 
              & $-25$ & MN & $800+3\times2700$             & 2.5 \\ 
NGC~7556      & $-64$ & MJ & $600+4\times2700$             & 3.2 \\ 
              & $-19$ & DG & $600+3\times2700$             & 2.4 \\ 
              & $+26$ & MN & $600+4\times2700$             & 3.2 \\ 
\noalign{\smallskip}       
\hline
\end{tabular} 
\end{small}
\end{center}     
\tablefoot{(1) Galaxy name. (2) Slit position angle measured North
  through East. (3) Slit position: MJ = major axis, MN = minor axis,
  DG = diagonal axis. (4) Number and exposure time of the single
  exposures. (5) Total exposure time.}
\end{table}    
%%%%%%%%%%%%%%%%%%%%%%%%%%%%%%%%%%%%%%%%%%%%%%%%%%%%%%%%%%%%%%%%%%%%  

We reduced the spectra using standard tasks in IRAF, as done in
\citet{Corsini2017}. The reduction steps included the subtraction of
bias, correction for internal and sky flat-field, trimming of the
spectra, identification and removal of bad pixels and cosmic rays,
correction for CCD misalignment, subtraction of the sky contribution,
and wavelength and flux calibration. Finally, we aligned and coadded
the spectra obtained for the same galaxy along the same axis using as
reference the center of the galaxy light profile extracted along the
spatial direction.

%%%%%%%%%%%%%%%%%%%%%%%%%%%%%%%%%%%%%%%%%%%%%%%%%%%%%%%%%%%%%%%%%%%% 
% FIGURE: SPECTRA
\begin{figure}[t!]
\centering
\includegraphics[width=0.49\textwidth]{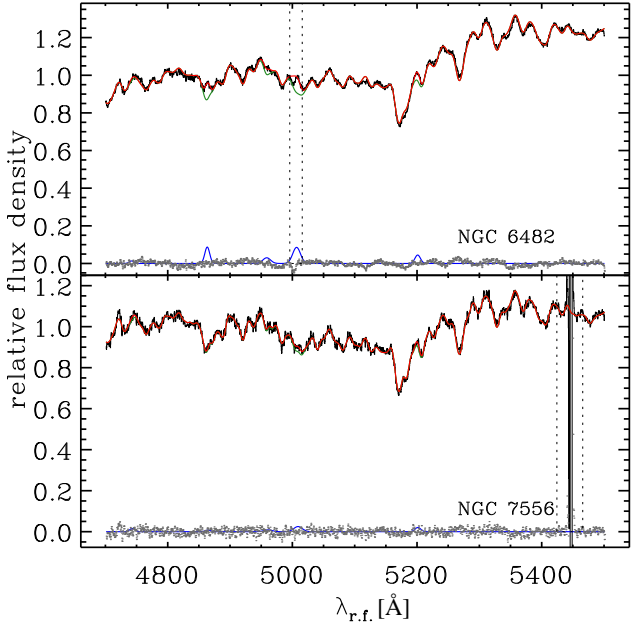}
\caption{Central rest-frame spectra extracted along the major axis of
  NGC~6482 (top panel) and NGC~7556 (bottom panel). Relative fluxes
  have false zero points for viewing convenience. In each panel the
  best-fitting model (red line) is the sum of the spectra of the
  ionized-gas (blue line) and stellar component (green line). The
  latter is obtained convolving the synthetic templates with the
  best-fitting LOSVD and multiplying them by the best-fitting Legendre
  polynomials. The residuals (gray dots) are obtained by subtracting
  the model from the spectrum. The vertical lines mark the wavelength
  ranges excluded from the fit.}
\label{fig:spectra}
\end{figure}
%%%%%%%%%%%%%%%%%%%%%%%%%%%%%%%%%%%%%%%%%%%%%%%%%%%%%%%%%%%%%%%%%%%% 

\subsection{Stellar kinematics}
\label{sec:kinematics}

We measured the stellar kinematics of both the galaxies with the
Penalized Pixel Fitting \citep[PPXF,][]{Cappellari2004} algorithm,
which we adapted to deal with the OSIRIS spectra.

%%%%%%%%%%%%%%%%%%%%%%%%%%%%%%%%%%%%%%%%%%%%%%%%%%%%%%%%%%%%%%%%%%%% 
% FIGURE: KINEMATICS AND LINE-STRENGTH INDICES N6482
\begin{figure*}[t!]
\centering
\includegraphics[angle=90.0,width=0.46\textwidth]{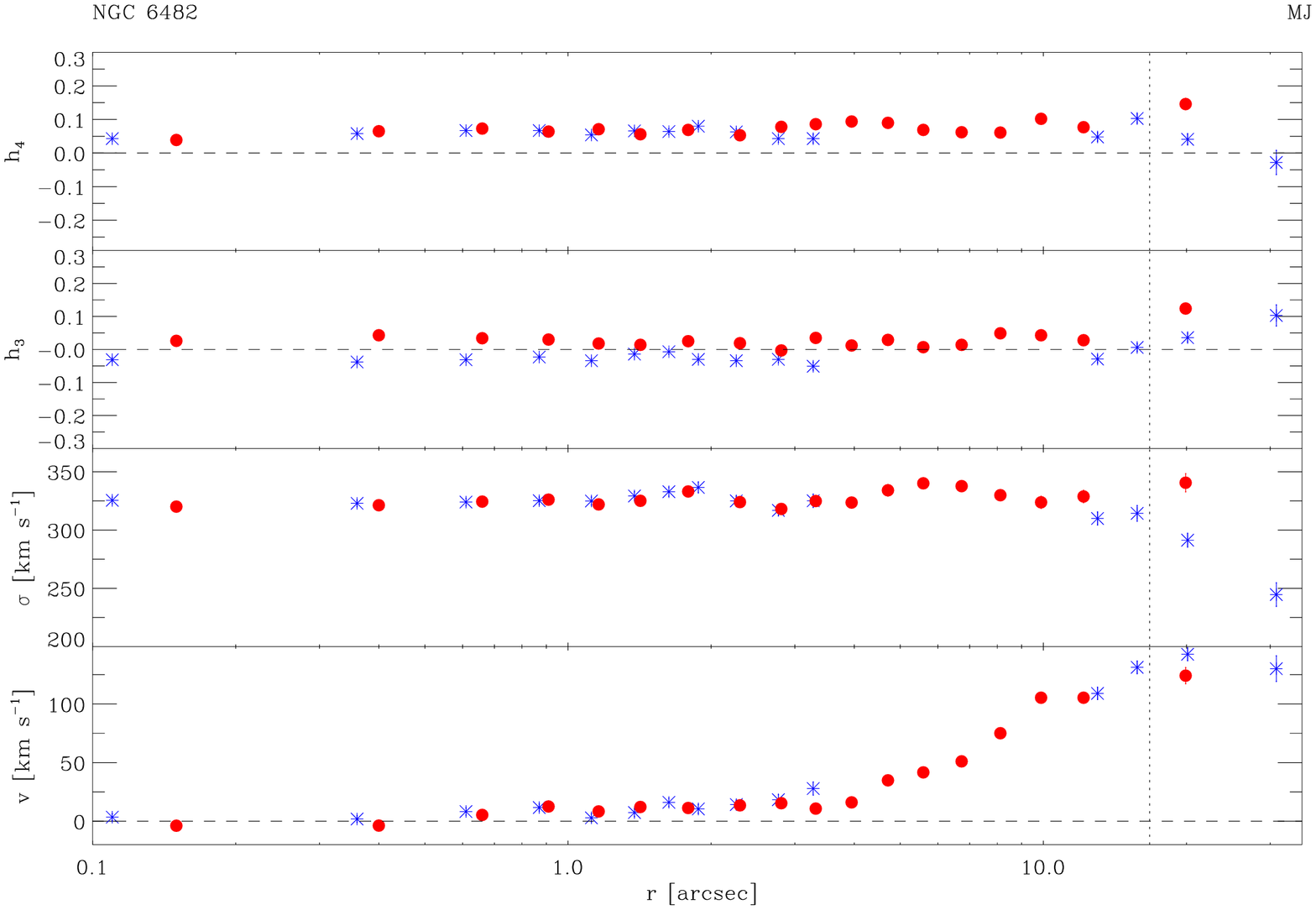}
\includegraphics[angle=90.0,width=0.46\textwidth]{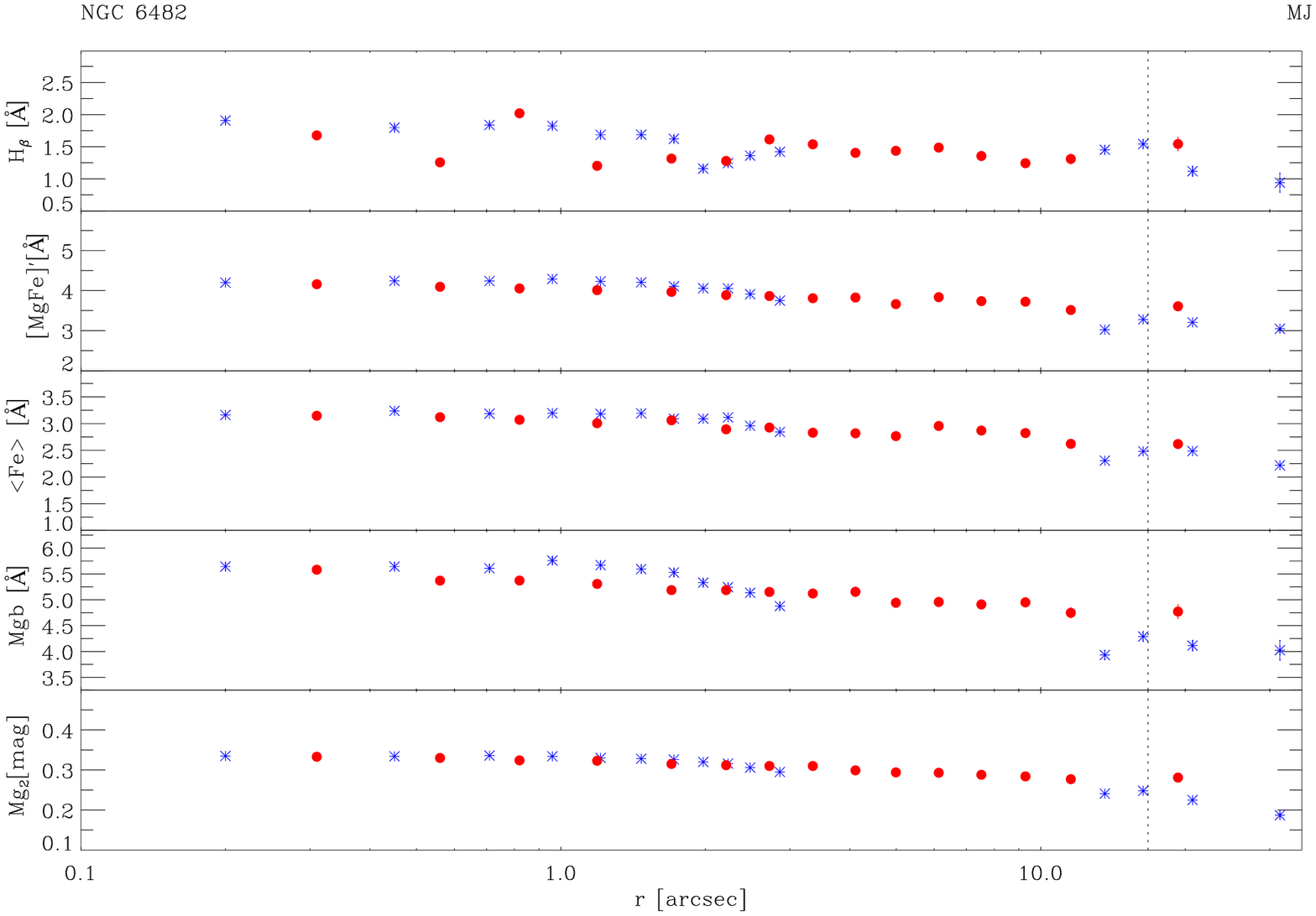}
\includegraphics[angle=90.0,width=0.46\textwidth]{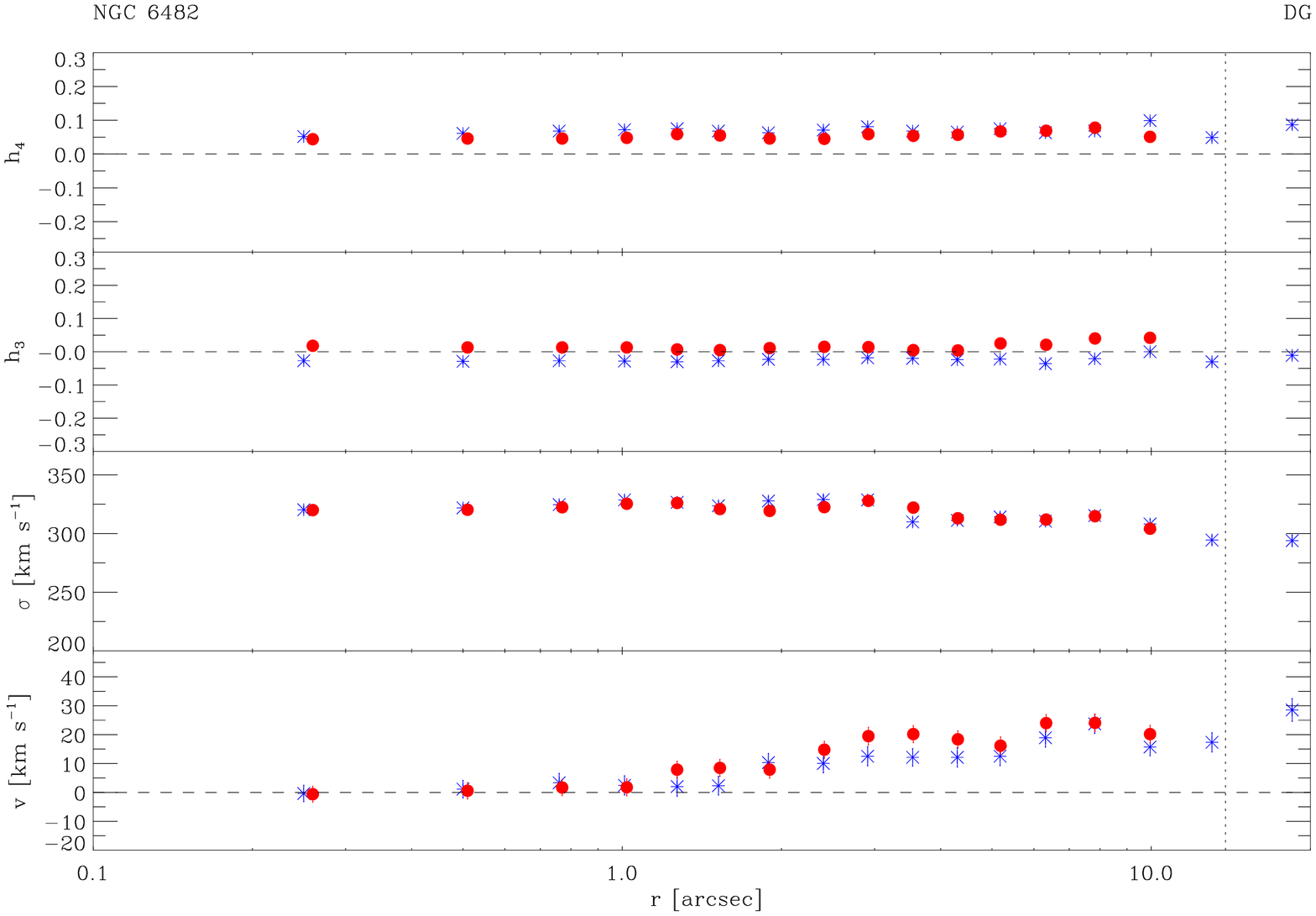}
\includegraphics[angle=90.0,width=0.46\textwidth]{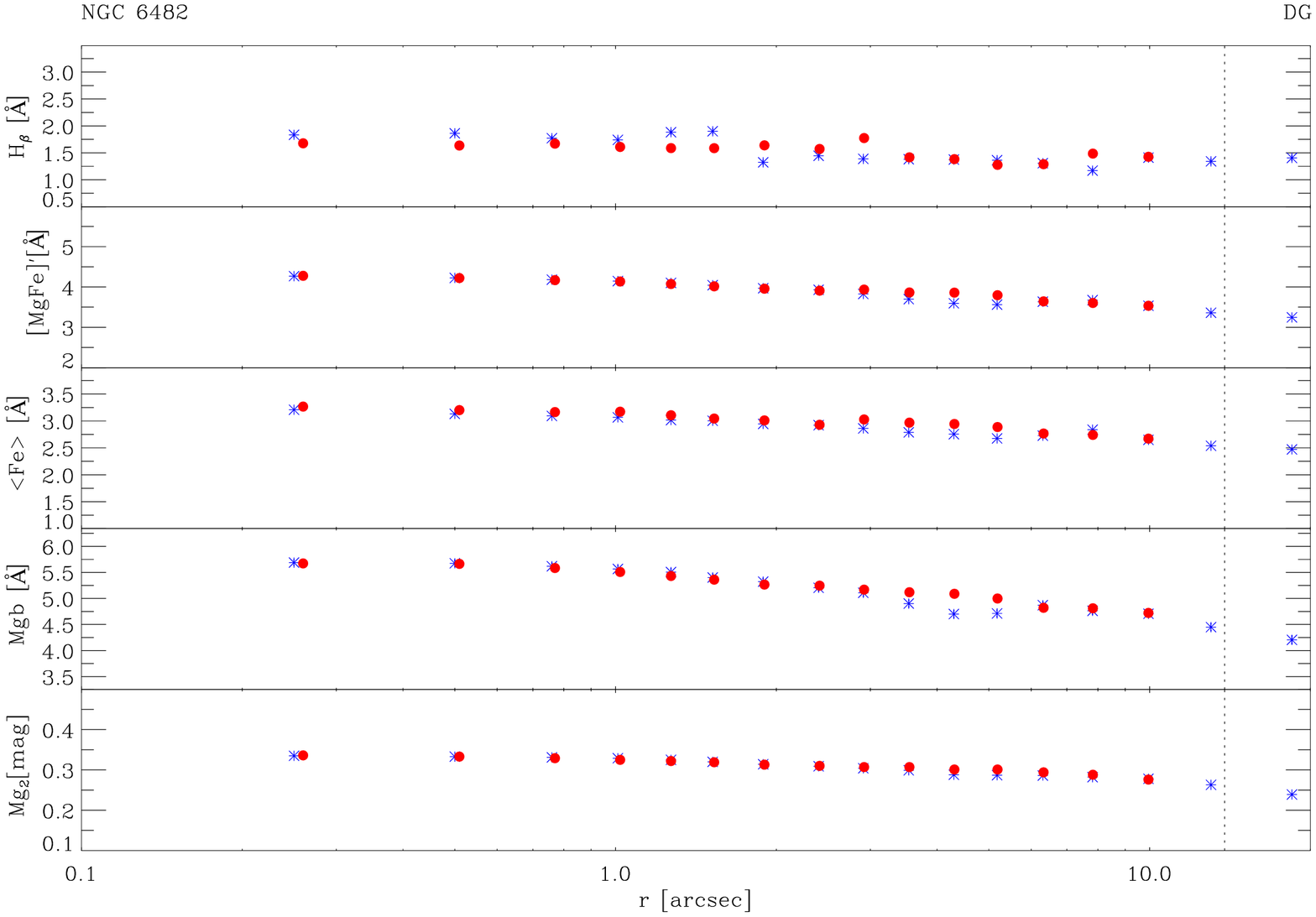}
\includegraphics[angle=90.0,width=0.46\textwidth]{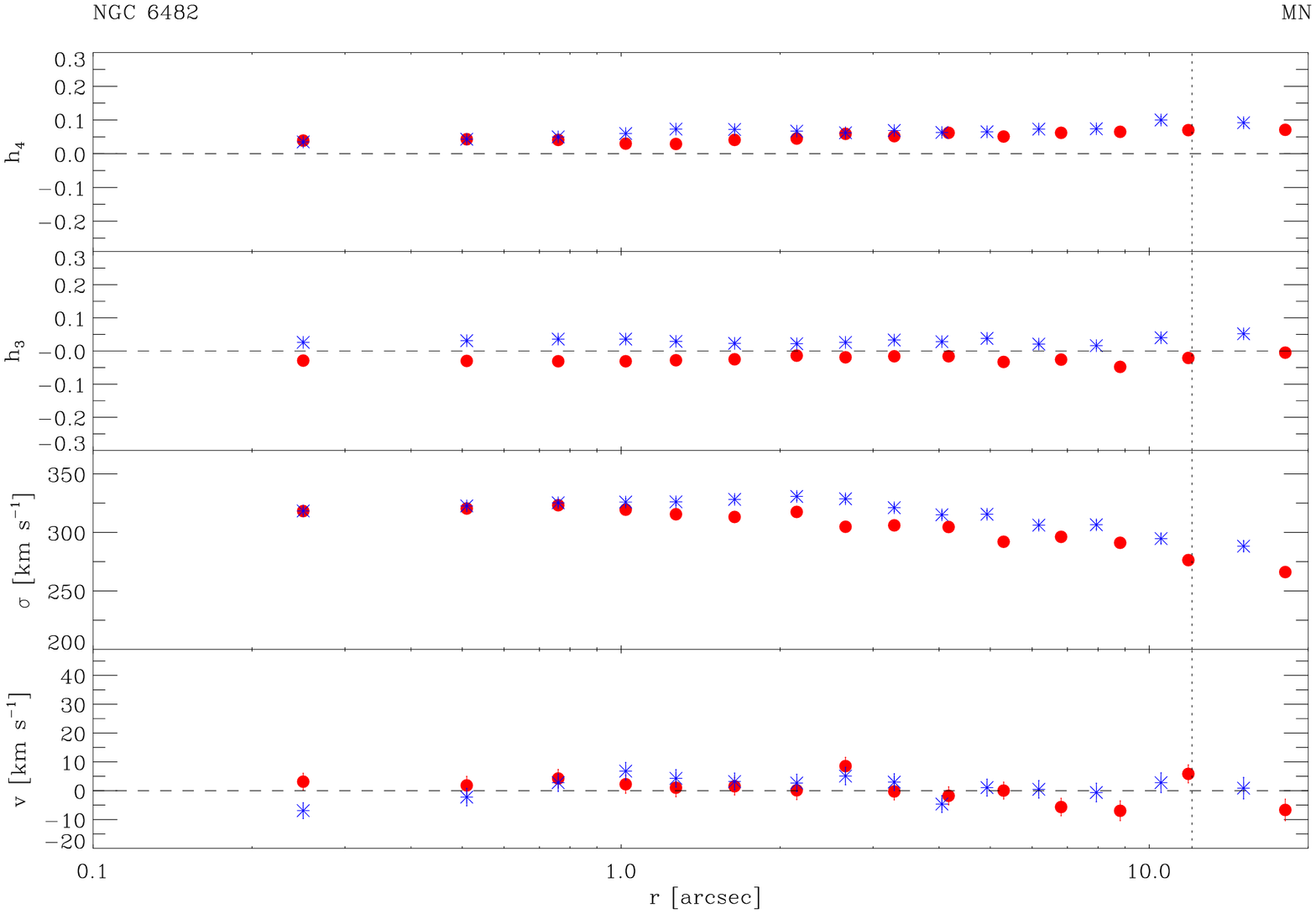}
\includegraphics[angle=90.0,width=0.46\textwidth]{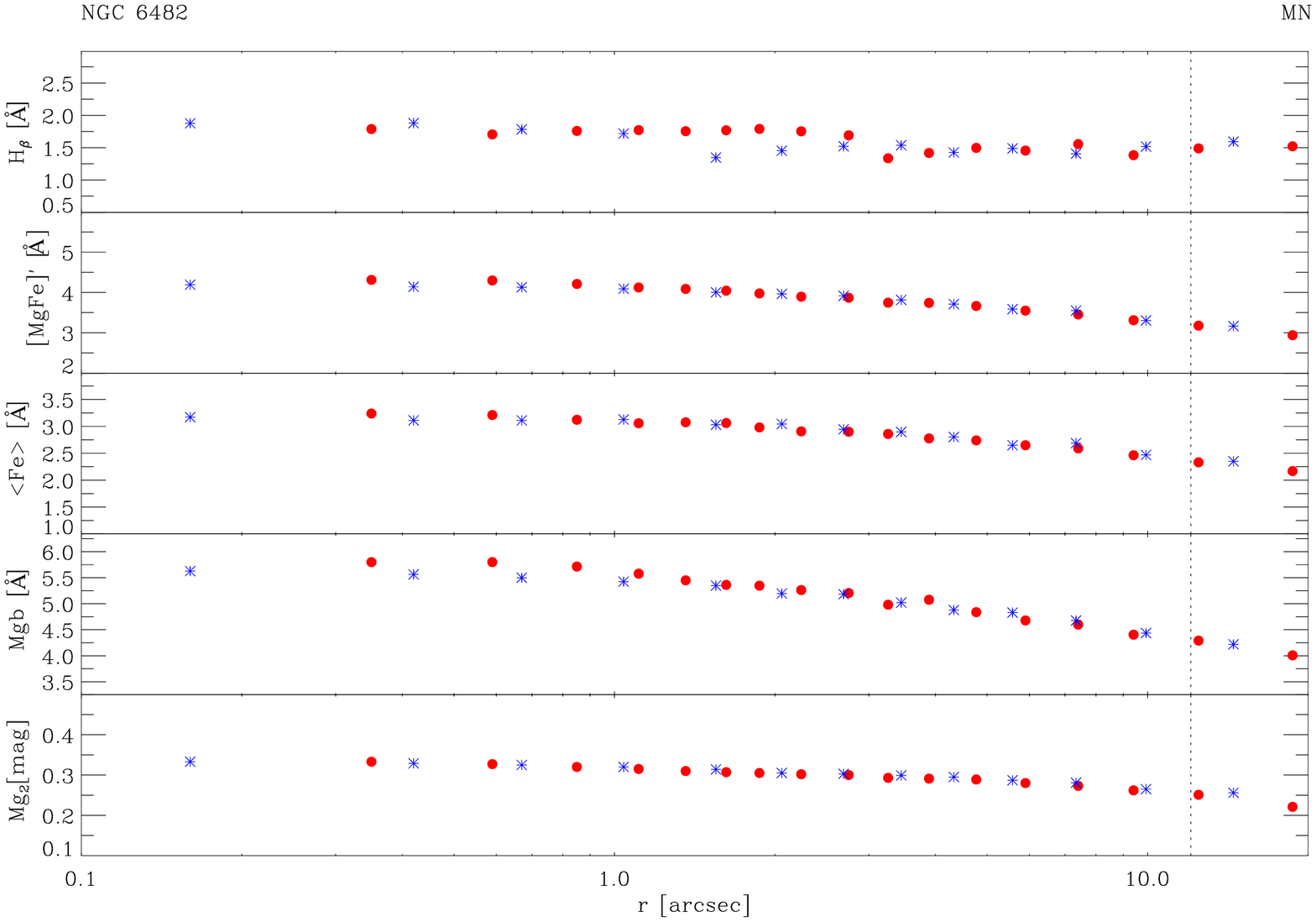}
\caption{Stellar kinematics and line-strength indices measured along
  the major (top panels), diagonal (middle panels) and minor axis
  (bottom panels) of NGC~6482. For each axis, the curves are folded
  around the nucleus. Blue asterisks and red circles refer to data
  measured along the approaching and residing sides of the galaxy,
  respectively.  The left panels (from top to bottom) show the radial
  profiles of the mean velocity ($v$) after the subtraction of the
  systemic velocity, velocity dispersion ($\sigma$), third- and
  fourth-order coefficients of the Gauss-Hermite decomposition of the
  LOSVD ($h_3$ and $h_4$).  The right panels (from top to bottom) show
  the radial profiles of the line-strength indices \Hb, \MgFe, \Fe,
  \Mgb, and \Mgd .  The vertical dotted line corresponds to the
  effective radius $r_{\rm e}$ of the spheroidal component.  The name
  of the galaxy and orientation of the slit are given for each data
  set.}
\label{fig:kinematics_indices_n6482}
\end{figure*}
%%%%%%%%%%%%%%%%%%%%%%%%%%%%%%%%%%%%%%%%%%%%%%%%%%%%%%%%%%%%%%%%%%%% 

%%%%%%%%%%%%%%%%%%%%%%%%%%%%%%%%%%%%%%%%%%%%%%%%%%%%%%%%%%%%%%%%%%%% 
% FIGURE: KINEMATICS AND LINE-STRENGTH INDICES N7556
%\addtocounter{figure}{-1}
\begin{figure*}[t!]
\centering
\includegraphics[angle=90.0,width=0.46\textwidth]{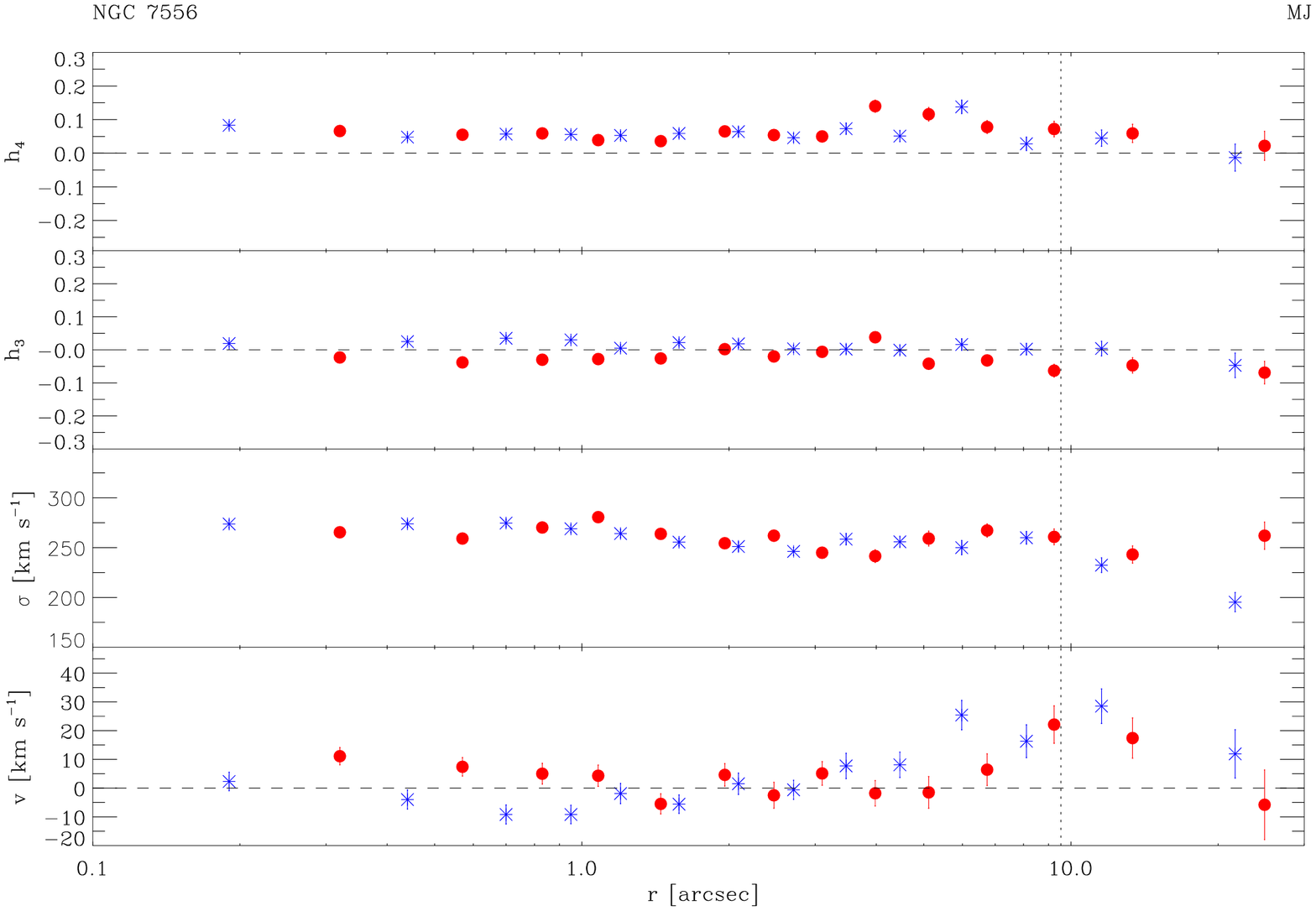}
\includegraphics[angle=90.0,width=0.46\textwidth]{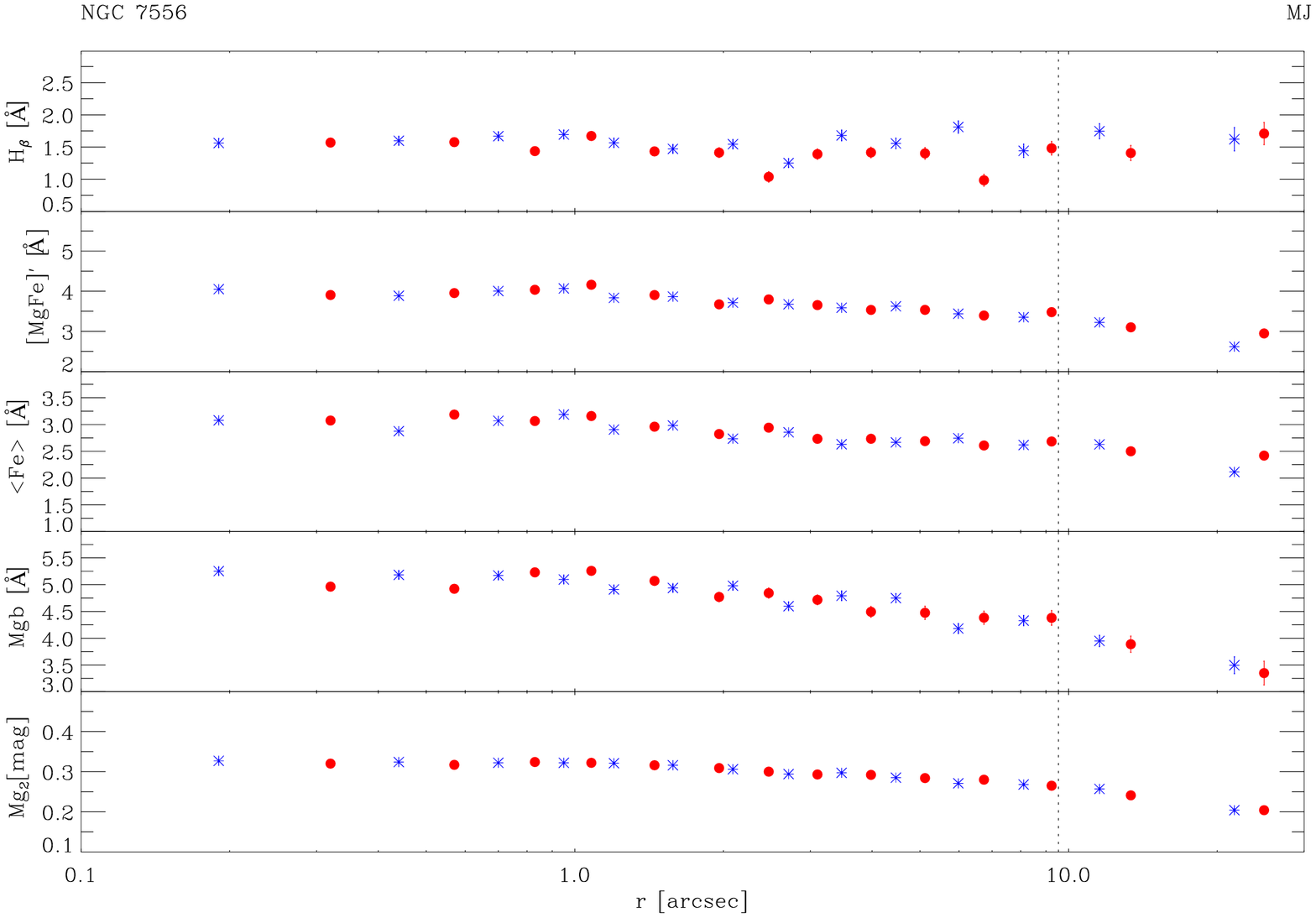}
\includegraphics[angle=90.0,width=0.46\textwidth]{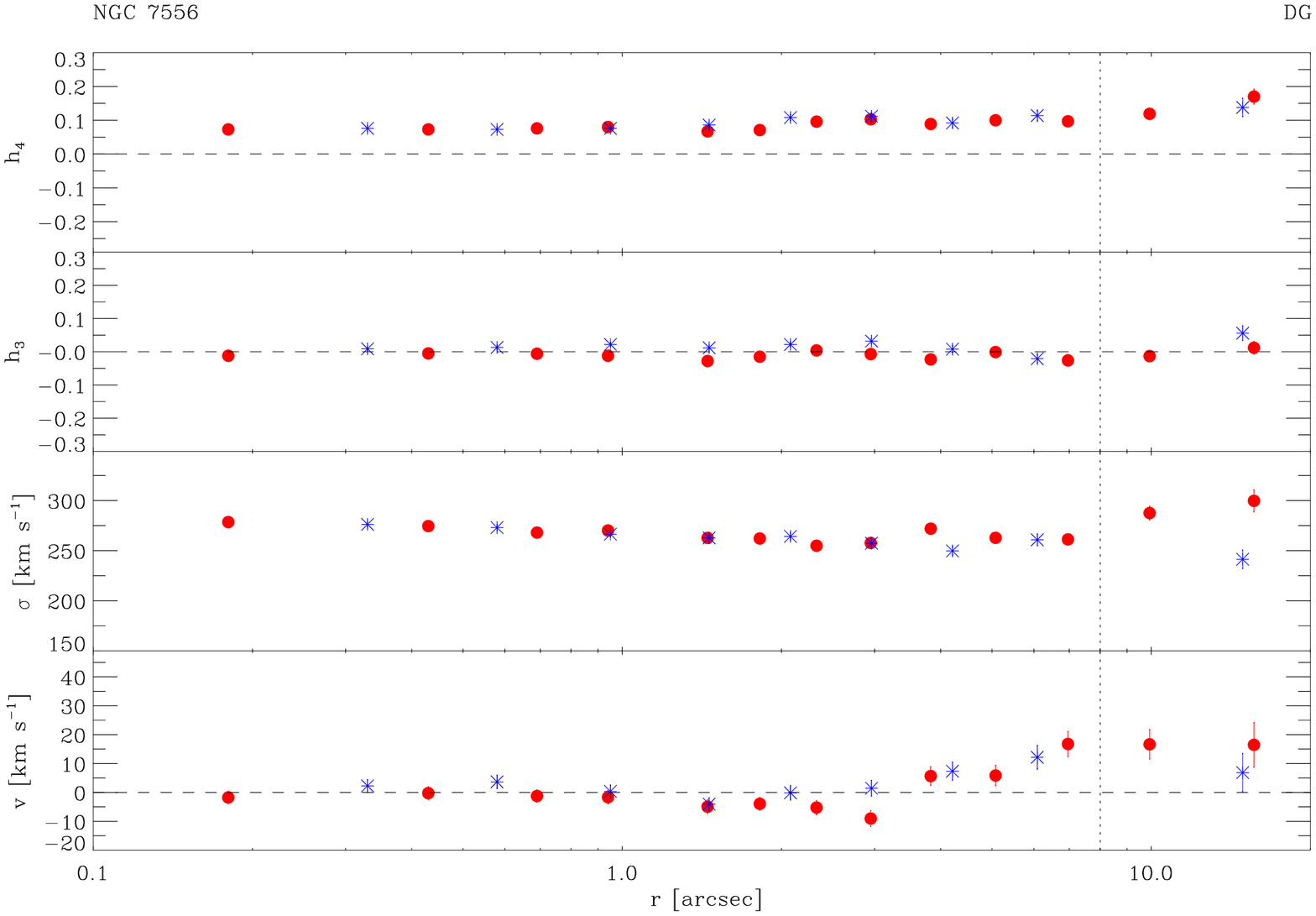}
\includegraphics[angle=90.0,width=0.46\textwidth]{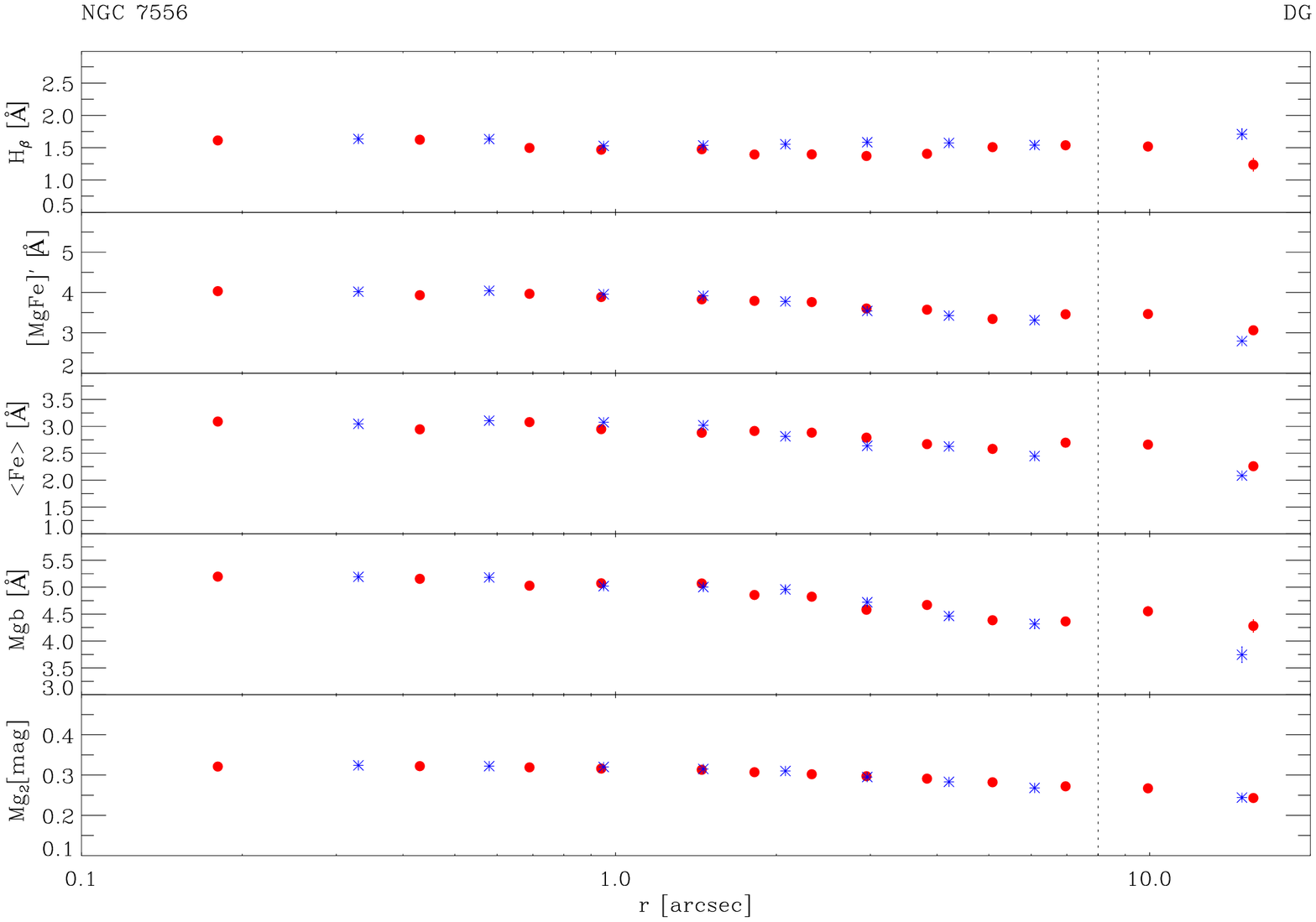}
\includegraphics[angle=90.0,width=0.46\textwidth]{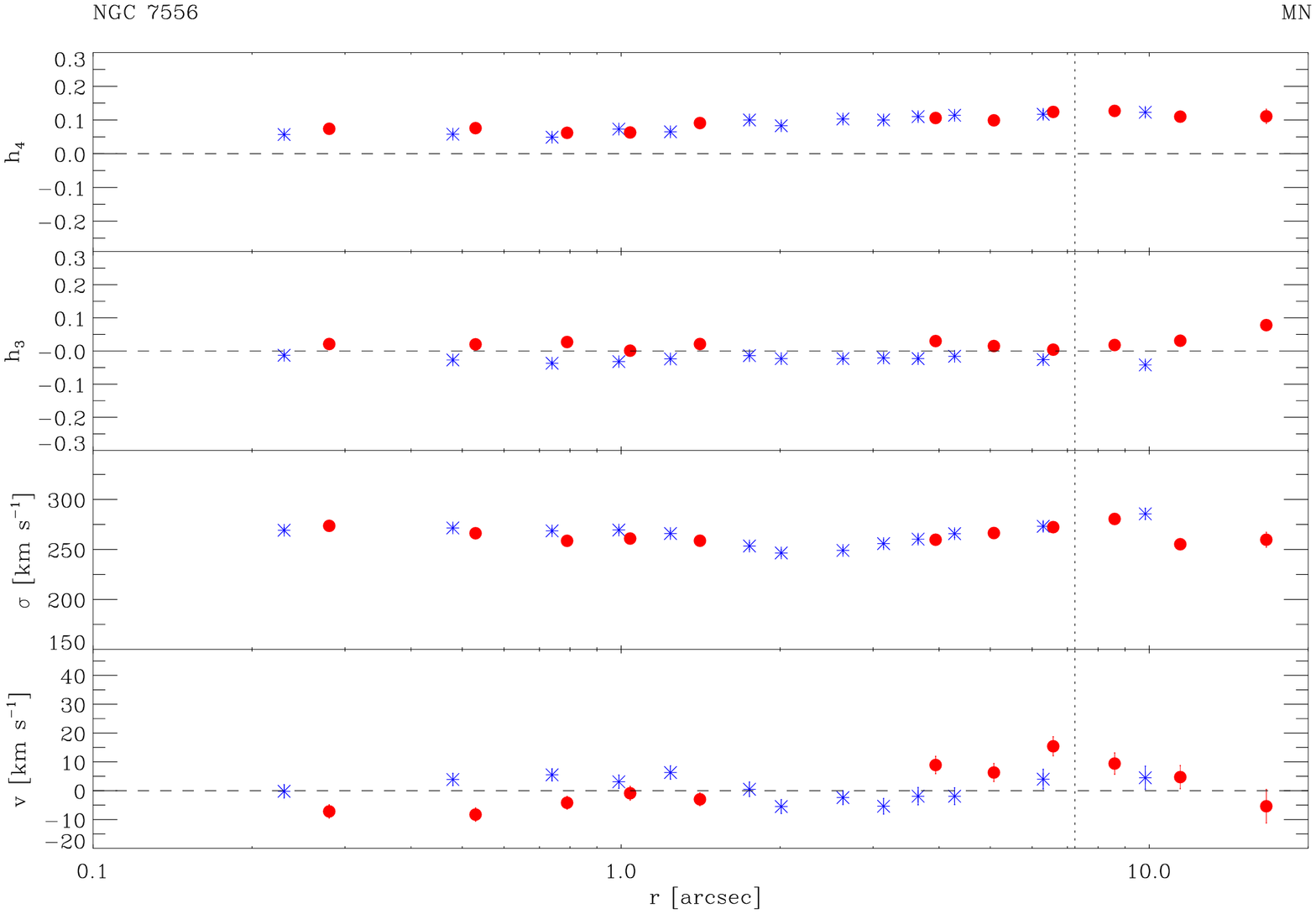}
\includegraphics[angle=90.0,width=0.46\textwidth]{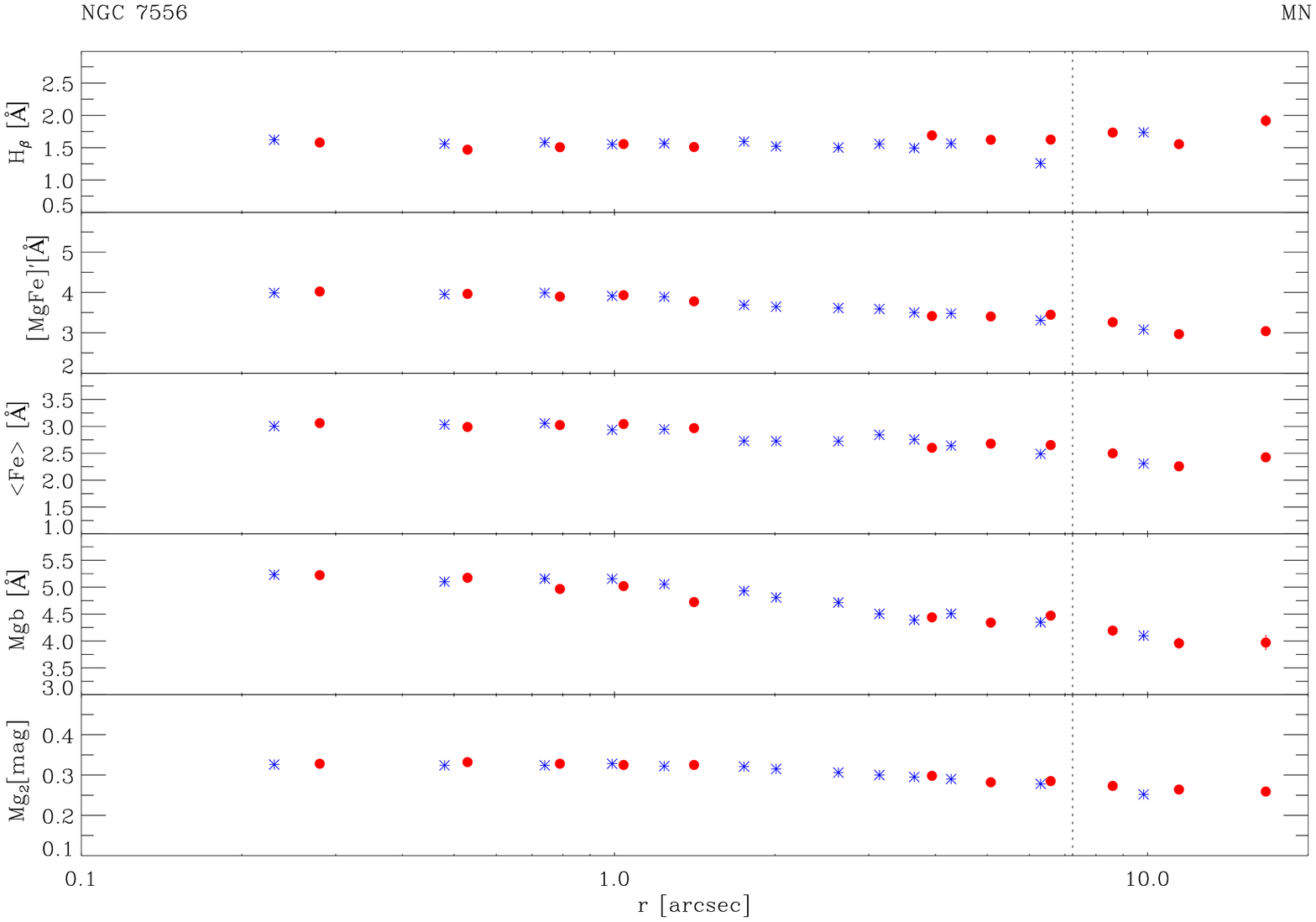}
\caption{Same as in Fig.~\ref{fig:kinematics_indices_n6482} but for
  NGC~7556.}
\label{fig:kinematics_indices_n7556}
\end{figure*}
%%%%%%%%%%%%%%%%%%%%%%%%%%%%%%%%%%%%%%%%%%%%%%%%%%%%%%%%%%%%%%%%%%%% 

We rebinned each summed galaxy spectrum along the dispersion direction
to a logarithmic scale. Then, we averaged the rebinned spectrum along
the spatial direction to obtain a signal-to-noise ratio $S/N \geq 70$
per resolution element. The resulting spectra are characterized by a
maximum $S/N = 327$ per resolution element in the innermost radial bin
along the major axis of NGC~6482 and a minimum $S/N = 72$ per
resolution element in the outermost radial bins along the major axis
of NGC~7556. In Fig.~\ref{fig:spectra} we show, as an example, the
central rest-frame spectra extracted along the major axis of NGC~6482
and NGC~7556. For each radial bin, we convolved a linear combination
of stellar spectra from the Medium Resolution Isaac Newton Telescope
Library of Empirical Spectra \citep[MILES,][]{SanchezBlazquez2006a,
  FalconBarroso2011} with a line-of-sight velocity distribution
(LOSVD) modeled as a truncated Gauss-Hermite series \citep{Gerhard1993,
  vanderMarel1993}
\begin{equation}
L(v_{\rm los})=\frac{e^{-\frac{1}{2}w^2}}{\sqrt{2\pi}} \left[ 1 + 
  \frac{h_3}{\sqrt{3}} \left( 2w^3 -3w \right) + 
  \frac{h_4}{\sqrt{6}} \left( 2w^4 -6w^2 +\frac{3}{2} \right) \right],
\end{equation}
where $w=(v_{\rm los}-v)/\sigma$ to fit the galaxy spectrum by a
$\chi^2$ minimization in pixel space.
To this aim, we degraded the spectral resolution of the stellar
spectra \citep[FWHM = 2.54 \AA,][]{Beifiori2011} by convolving them
with a Gaussian function in order to match the spectral resolution of
the galaxy spectra. After rebinning the stellar spectra to a
logarithmic scale along the dispersion direction, we de-reshifted them
to rest frame and cropped their wavelength range to match the
redshifted frame of the galaxy spectra.
Moreover, we added a low-order multiplicative Legendre polynomial
to correct for reddening and large-scale residuals of
flat-fielding. We excluded from the fitting procedure the wavelength
ranges with a spurious signal coming from the imperfect subtraction of
cosmic rays and bright sky emission lines.

In this way, we determined the value of the mean velocity $v$,
velocity dispersion $\sigma$, and the third- and fourth-order LOSVD
moments $h_3$ and $h_4$ of the stellar component along the different
observed axes of both galaxies. We assumed the statistical errors on
the stellar kinematic parameters to be the formal errors of the pPXF
best fit after rescaling the minimum $\chi^2$ to achieve $\chi^2_{\rm
  min}=N_{\rm dof}=N_{\rm d}-N_{\rm fp}$, with $N_{\rm dof}$, $N_{\rm
  d}$, and $N_{\rm fp}$ the number of the degrees of freedom, data
points, and fitting parameters, respectively \citep{Press1992}. We
report the measured stellar kinematics of both galaxies in
Table~\ref{tab:kinematics}.

We fitted the emission lines present in the galaxy spectra with
Gaussian functions using the Gas and Absorption Line Fitting
\citep[GANDALF,][]{Sarzi2006} algorithm. The \oiiiqc\ and
\niqc\ emission doublets were nearly detected in the spectra of
NGC~6482. Indeed, they have $S/rN\gtrsim3$, where we estimated the
residual noise $rN$ as the standard deviation of the difference
between the galaxy and best-fitting stellar spectrum. No emission line
was detected in NGC~7556 (Fig.~\ref{fig:spectra}).

We show the folded kinematic profiles of NGC~6482 and NGC~7556 in
Figs.~\ref{fig:kinematics_indices_n6482} and
\ref{fig:kinematics_indices_n7556}. We plot the stellar velocities
with respect to the galaxy center after subtracting the systemic
velocity and without correcting for galaxy inclination, while the
velocity dispersions are corrected for instrumental FWHM. There is a
good agreement between the radial profiles of the kinematic parameters
measured on either side of each axis and between the different axes of
each galaxy.

Along the major axis, NGC~6482 is slowly rotating for $r \la 3$
arcsec, where the velocity linearly increases to $v \simeq 20$ \kms
. Outwards, the velocity sharply rises to a maximum of $v \simeq 110$
\kms\ at $r \simeq 10$ arcsec and then it stays constant out to the
last measured radius. A velocity of $v \simeq 20$ \kms\ is observed
along the diagonal axis for $3 \la r \la 18$ arcsec, while no rotation
is measured along the minor axis. The velocity dispersion profile is
almost flat at $\sigma \simeq 320$ \kms\ within the innermost 3 arcsec
and it mildly declines to $\sigma \simeq 250$ \kms\ at larger
radii. Constant $h_3 \simeq 0$ and $h_4 \simeq 0.05$ are measured.
The measured kinematics is in agreement with the integral-field
spectroscopic data by \citet{Raskutti2014}, who investigated the
kinematic properties of the stellar halos of a sample of massive 
ETGs \citep{Greene2013}.

NGC~7556 displays a velocity $v \la 20$ \kms\ for $r \ga 5$ arcsec
along all the observed axes. No rotation is measured for $r>r_{\rm
  se}$ confirming that the stellar envelope is not a rotating disk.
The velocity dispersion drops from a central value of $\sigma \simeq
270$ \kms\ to about $250$ \kms\ at $r \simeq 2$ arcsec and it remains
nearly constant further out. The Gauss-Hermite coefficients are
characterized by a flat radial profile of $h_3 \simeq 0$ and a profile
of $h_4$ rising from about 0.05 to 0.1 within 2 arcsec and flattening
outwards. The measured kinematics is consistent with that obtained
with integral-field spectroscopy by \citet{Veale2017} for the 
MASSIVE Survey, which targets the most massive ETGs in the local
universe \citep{Ma2014}.

%%%%%%%%%%%%%%%%%%%%%%%%%%%%%%%%%%%%%%%%%%%%%%%%%%%%%%%%%%%%%%%%%%%% 
% FIGURE: POPULATION GRID FOR N6482
\begin{figure*}[t!]
\centering
\includegraphics[width=0.49\textwidth]{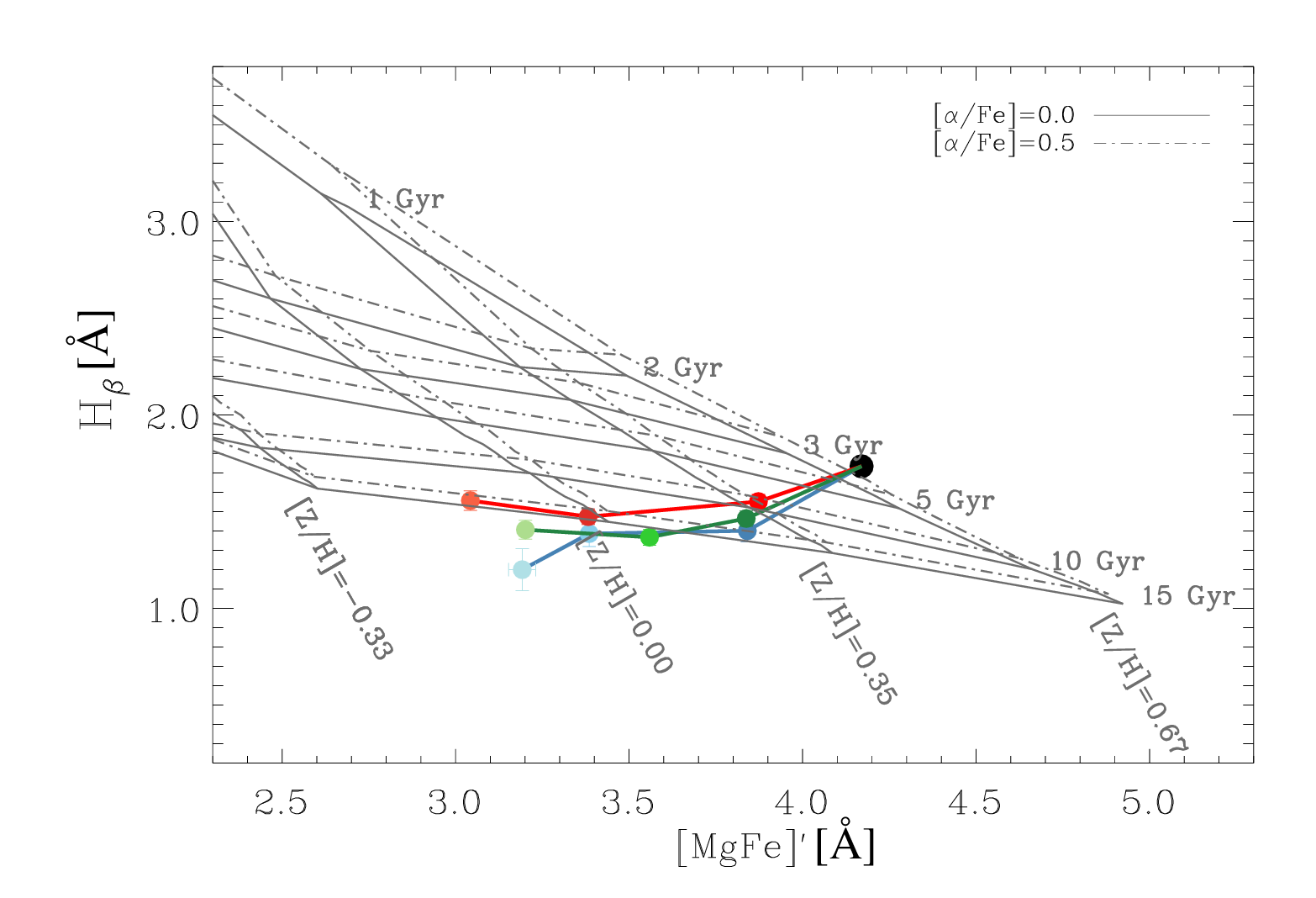}
\includegraphics[width=0.49\textwidth]{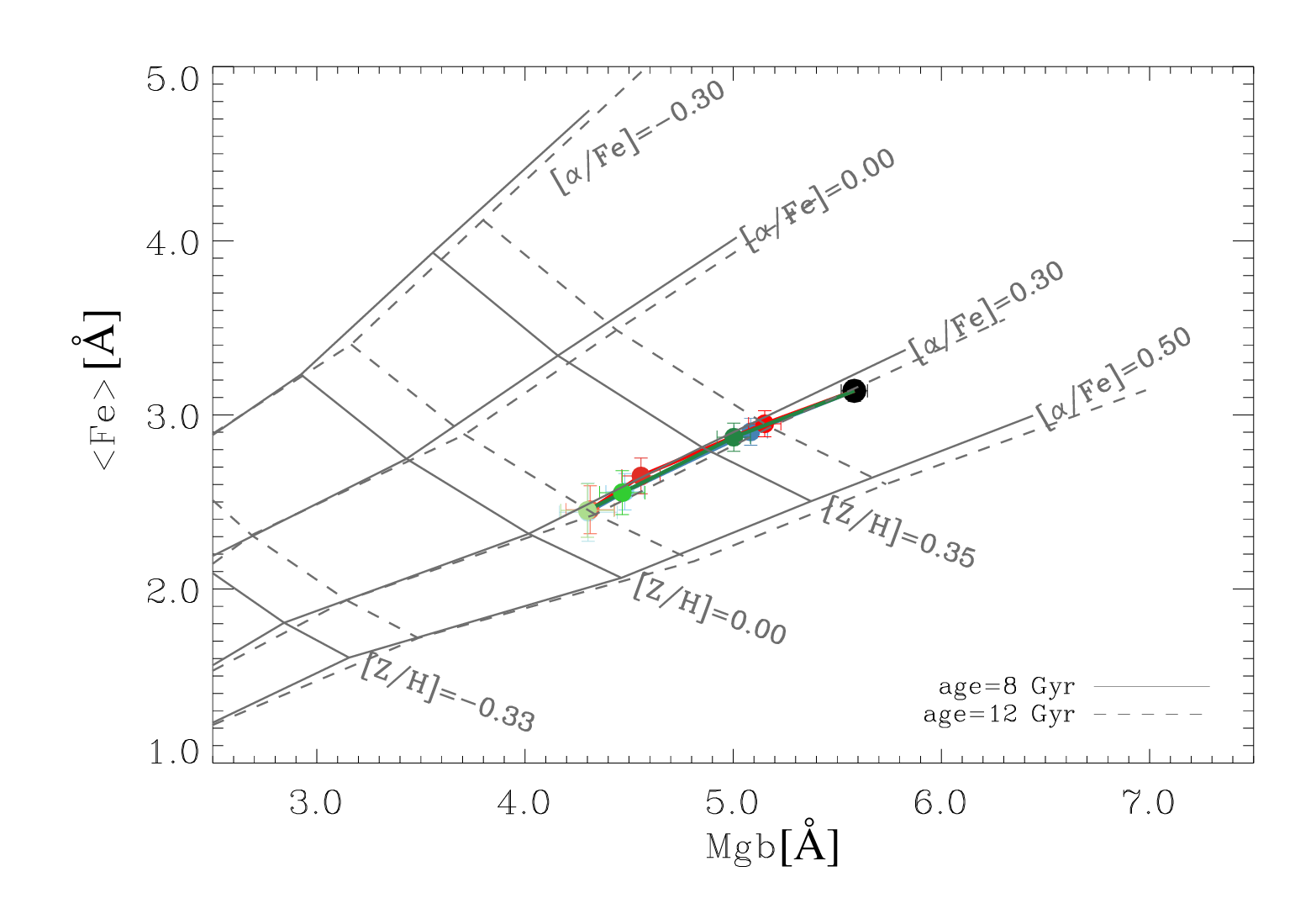}
\caption{Averaged values of the \Hb\ and \MgFe\ (left panel) and
  \Fe\ and \Mgb\ line-strength indices (right panel) measured inside
  an aperture of radius $r = 0.1 r_{\rm e}$ (black circles) of
  NGC~6482. The averaged line-strength indices measured for $0.1
  r_{\rm e} < r \leq 0.5 r_{\rm e}$, $0.5 r_{\rm e} < r \leq r_{\rm
    e}$, and $r > r_{\rm e}$ are plotted with circles of
  different color tone (from darker to lighter), respectively. Blue,
  green, and red circles correspond to data obtained along the major,
  diagonal, and minor axes, respectively. The gray lines indicate the models
  by \citet{Thomas2003}. In the left panel, the age-metallicity grids
  are plotted with two different $\alpha/$Fe enhancements: \aFe\ = 0
  dex (continuous lines) and \aFe\ = 0.5 dex (dash-dotted lines). In the
  right panel, the \aFe -metallicity grids are plotted with two
  different ages: 2 Gyr (continuous lines) and 8 Gyr (dash-dotted lines).}
\label{fig:pops_n6482}
\end{figure*}
%%%%%%%%%%%%%%%%%%%%%%%%%%%%%%%%%%%%%%%%%%%%%%%%%%%%%%%%%%%%%%%%%%%% 

%%%%%%%%%%%%%%%%%%%%%%%%%%%%%%%%%%%%%%%%%%%%%%%%%%%%%%%%%%%%%%%%%%%% 
% FIGURE: POPULATION GRID FOR N7556
\begin{figure*}[t!]
\centering
\includegraphics[width=0.49\textwidth]{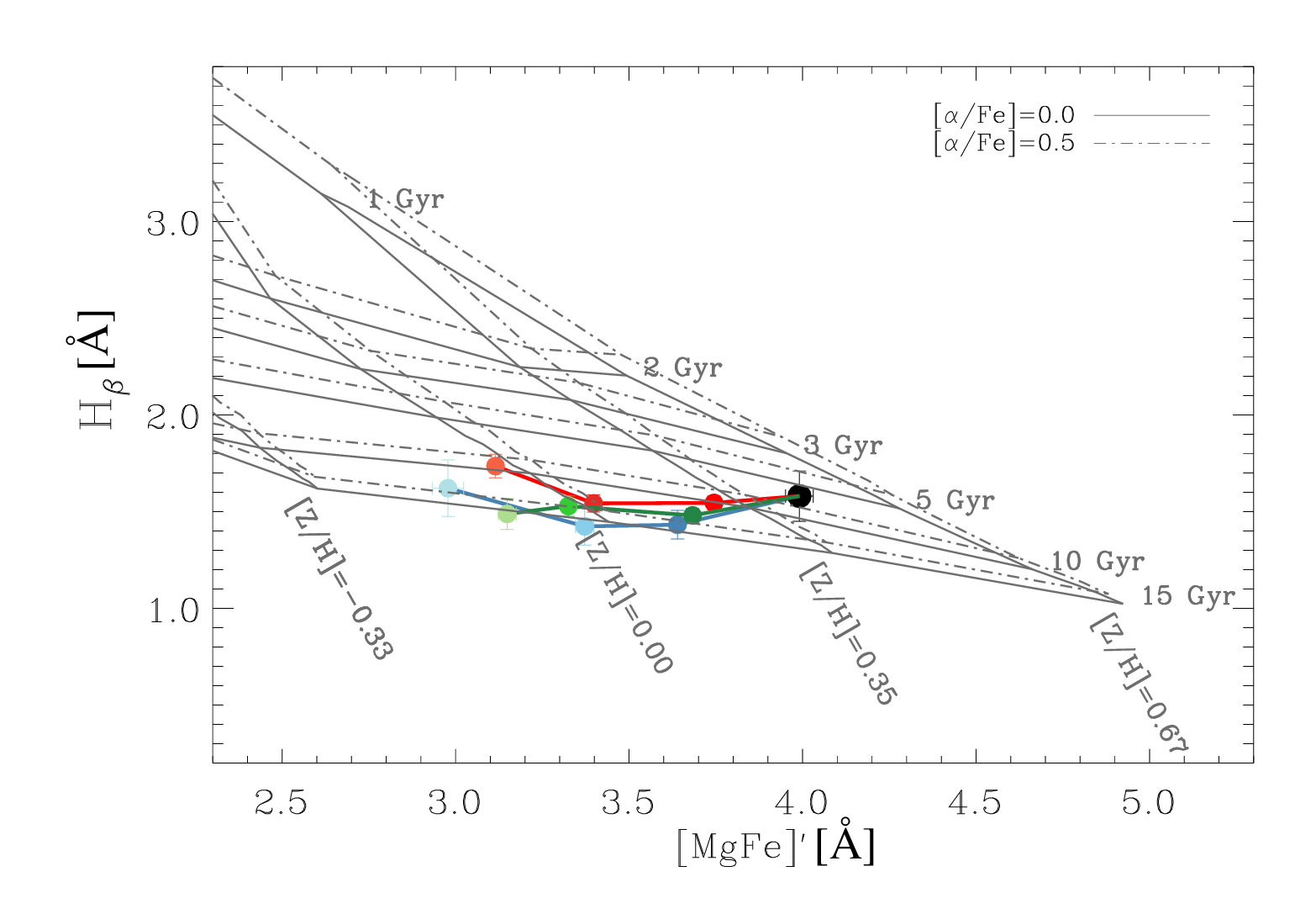}
\includegraphics[width=0.49\textwidth]{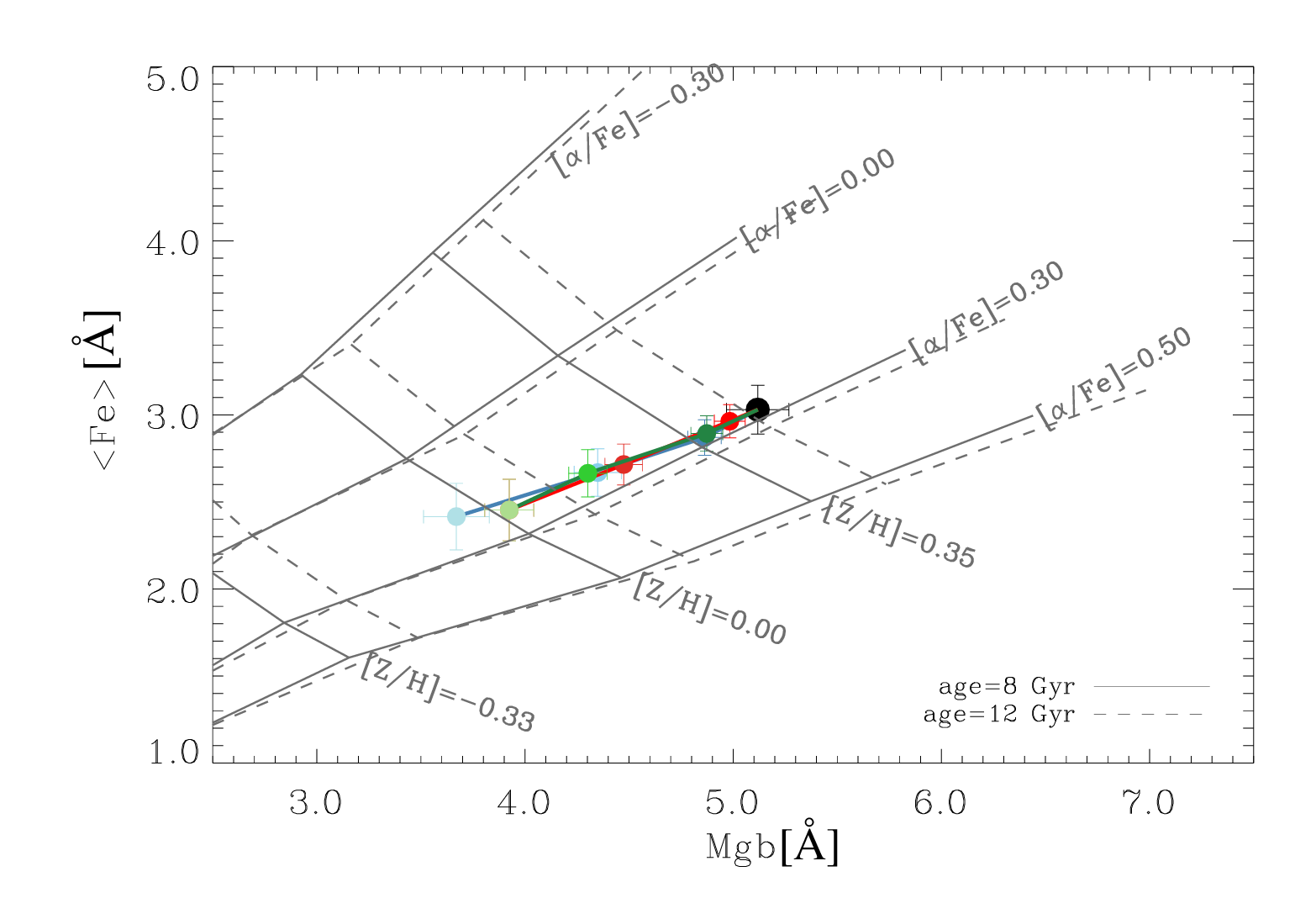}
\caption{Same as in Fig.~\ref{fig:pops_n6482} but for NGC~7556.}
\label{fig:pops_n7556}
\end{figure*}
%%%%%%%%%%%%%%%%%%%%%%%%%%%%%%%%%%%%%%%%%%%%%%%%%%%%%%%%%%%%%%%%%%%% 

\subsection{Line-strength indices}
\label{sec:indices}

We measured the Mg, Fe, and \Hb\ line-strength indices of the
Lick/Cassegrain Image Dissector Scanner Spectrograph (Lick/IDS) system
\citep{Worthey1994} from the galaxy spectra rebinned in the dispersion
and radial directions. Then, we computed the average iron index \Fe
$\rm = (Fe_{5270}+Fe_{5335})/2$ \citep{Gorgas1990} and the combined
magnesium-iron index \MgFe $= \sqrt{ {\rm Mg{\it b} (0.72 \times
    Fe_{5270}+0.28 \times Fe_{5335}) }}$ \citep{Thomas2003}. We took
into account the difference between our and Lick/IDS spectral
resolutions by degrading our spectra to match the Lick/IDS resolution
\citep[$\rm FWHM=8.6$ \AA ,][]{Worthey1997} before measuring the
line-strength indices. The offsets between our line-strength
measurements and Lick/IDS line-strength values were smaller than the
mean error of the differences and therefore we did not apply any
offset correction to our line-strength measurements. We also corrected
the line-strength indices for broadening due to the galaxy velocity
dispersion as in \citet{Trager1998}. The same method was adopted by
\citet{Eigenthaler2013} to correct the line-strength indices of their
FG BCGs.

We derived the errors on the line-strength indices by means of Monte
Carlo simulations, as done in \citet{Morelli2015}. At each radius, we
generated a set of simulated galaxy spectra by adding photon, readout,
and sky noise to the best-fitting stellar spectrum obtained from the
kinematic analysis. We measured the line-strength indices in the
simulated spectra as if they were real. We used the mean and standard
deviation of the line-strength indices in the simulated spectra to
estimate the systematic $\sigma_{\rm syst}$ and statistical
$\sigma_{\rm stat}$ errors of the measured indices, respectively. We
computed the errors as $\sigma^2$ = ${\sigma_{\rm stat}^2 +
  \sigma_{\rm syst}^2}$.

We list the measured values of \Hb , \Mgd, \Mgb , Fe$_{5270}$, and
Fe$_{5335}$ in Table~\ref{tab:indices} and plot the folded radial
profiles of \Hb, \Mgb, \MgFe , and \Fe\ in
Figs.~\ref{fig:kinematics_indices_n6482} and
\ref{fig:kinematics_indices_n7556}. 
For NGC~6482, there are significant differences in \Hb\ and
\Mgb\ (and consequently on \MgFe) on the two galaxy sides only within
the inner 3 arcsec of the major and minor axes. For NGC~7556 there is
a good agreement between the radial profiles of the measured
line-strength indices on either side of each axis. The central values
of the line-strength indices measured on the different axes are
consistent with each other for both galaxies.

All the radial profiles of the line-strength indices of NGC~6482 show
a central maximum followed by a mild decline to the outermost measured
radius with a gradient change observed at $r \simeq 3$ arcsec. We find
that \Hb\ decreases from about 1.8 to 1.5 \AA , \MgFe\ from about 4.3
to 3.0 \AA , \Fe\ from about 3.2 to 2.7 \AA , \Mgb\ from about 5.7 to
4.0 \AA , and \Mgd\ from about 0.35 to 0.20 mag. All the line-strength
indices we measured are consistent with those obtained by
\citet{Trager1998} with an aperture of $1.4\times4.0$ arcsec$^2$,
except for \Hb . This discrepancy is due to the different correction
adopted to deal with the contamination of the \Hb\ emission line,
which fills in the absorption line and lowers its actual equivalent
width. We measured the \Hb\ index on the best-fitting stellar spectrum
obtained from the kinematic analysis.
 
In NGC~7556, the \Hb\ index is almost constant at all radii, contrary
to the other line-strength indices. They are characterized by
decreasing radial profiles, which become flat along the minor and
diagonal axes for $r\ga5$ arcsec. The measurement for \Hb\ is
\Hb$\,\simeq 1.7$ \AA , whereas \MgFe\ decreases from about 4.0 to 3.0
\AA , \Fe\ from about 3.0 to 2.8 \AA , \Mgb\ from about 5.2 to 4.0 \AA
, and \Mgd\ from about 0.33 to 0.25 mag. The central value of \Mgd\ is
in agreement with that reported by \citet{Wegner2003} for a
circularized aperture of 0.9 arcsec at the adopted distance.

\section{Properties of the stellar populations}
\label{sec:populations}

We obtained the stellar population properties of NGC~6482 and NGC~7556
by comparing the measurements of the line-strength indices with the
model predictions by \citet{Thomas2003} for the single stellar
population as a function of age, metallicity, and $\alpha/$Fe
enhancement following \citet{Morelli2008, Morelli2012, Morelli2015}.
The central values of velocity dispersion $\sigma_0$ and line-strength
indices \Mgb, \Hb, \Fe, and \MgFe\ of both galaxies were derived from
the radial profiles along the major, diagonal, and minor axes. The data
points inside an aperture of radius $0.1 r_{\rm e}$ were averaged
adopting a relative weight proportional to their $S/N$. This is $r_{\rm
  e}=4.4$ and 4.6 kpc for NGC~6482 and NGC~7556, respectively.
We found that the central regions of NGC~6482 ($\sigma_0=323\,\pm\,3$
\kms) are made by relatively young stars (${\rm Age_0}=3.3\pm0.4$ Gyr)
with high metallicity (\ZH$_0\,>0.67$ dex) and supersolar $\alpha/$Fe
enhancement (\aFe$_0\,=0.38\pm 0.03$ dex).  This result holds also
when the central population properties are independently derived along
the three available axes, but it is different from
\citet{SanchezBlazquez2006b} who pointed out the presence of older
(${\rm Age_0}=11\pm1$ Gyr) and less metal rich (\ZH$_0\,=0.11\pm0.03$
dex) stars using spectral synthesis models. As a consequence, their
gradients of age and metallicity are much flatter than ours
\citep{SanchezBlazquez2006c}.
This difference in the population properties obtained for NGC~6482
could be due to the fact that \citet{SanchezBlazquez2006b} derived the
equivalent width of the \Hb\ emission line from that of \oiiic\ to
correct the \Hb\ line-strength index for the effect of emission. They
adopted the ratio {\it EW}(\Hb$_{\rm em}$)/{\it EW}(\oiiic)$\,=0.7$ by
\citet{Gonzalez1993}. However, there is evidence that this correction
is uncertain for individual galaxies \citep{Mehlert2000} although it
is good in a statistical sense \citep{Trager2000}.
The central stellar population of NGC~7556 ($\sigma_0=265\pm3$ \kms)
is older (${\rm Age_0}=7\pm1$ Gyr), less metal rich
(\ZH$_0\,=0.46\pm0.03$ dex), and with lower $\alpha/$Fe enhancement
(\aFe$_0\,=0.28\pm0.02$ dex) than NGC~6482.

For comparison, \citet{Khosroshahi2013} reported a few other cases of
FG BCGs hosting a secondary stellar component. It is on average less
old ($\langle {\rm Age_{\rm s}} \rangle \simeq10$ Gyr) and metal rich
($\langle$\ZH$_{\rm s}\rangle\,\simeq0.3$ dex) than the galaxy main
component ($\langle{\rm Age_{\rm m}}\rangle\simeq14$ Gyr,
$\langle$\ZH$_{\rm m}\rangle\,\simeq1.7$ dex). These secondary
components are in any case much older and more metal rich than that we
found in NGC~6482.

The radial trends of the line-strength indices \Mgb, \Hb, \Fe, and
\MgFe\ were derived along all the observed axes and are shown in
Fig.~\ref{fig:pops_n6482} for NGC~6482 and in
Fig.~\ref{fig:pops_n7556} for NGC~7556. We averaged the data points
along the major axis adopting a relative weight proportional to their
$S/N$ within $0.1 r_{\rm e} < r \leq 0.5 r_{\rm e}$, $0.5 r_{\rm e} <
r \leq r_{\rm e}$, and $r > r_{\rm e}$, respectively. We defined the
corresponding radial ranges along the diagonal and minor axes by
taking into account the misalignment with respect to the major
axis and axial ratio of the spheroidal component in order to map the
same galaxy regions along the different directions.
We did not find any significant difference in the properties of
stellar populations measured along the different axes of both NGC~6482
and NGC~7556, because the averaged line-strength indices measured on
the different axes at the same radius are nearly consistent within the
errors. As soon as the radius increases, the age and metallicity of
NGC~6482 steadily increases to ${\rm Age} \simeq 15$ Gyr and
decreases to \ZH$\,\simeq 0$ dex, respectively while the
$\alpha/$Fe enhancement is constant. Also the stellar population of
NGC~7556 is much older (${\rm Age} \simeq 15$ Gyr) at larger radii
with respect to the center, but it is characterized by a subsolar
metallicity (\ZH$\,\simeq -0.1$ dex) and lower $\alpha/$Fe enhancement
(\aFe$\,\simeq 0.2$ dex).

%%%%%%%%%%%%%%%%%%%%%%%%%%%%%%%%%%%%%%%%%%%%%%%%%%%%%%%%%%%%%%%%%%%% 
% TABLE: GRADIENTS
\renewcommand{\tabcolsep}{3pt}
\begin{table*}  
\caption{Stellar population gradients in the sample galaxies.
\label{tab:gradients}}
\begin{center}
\begin{small}
\begin{tabular}{lccccr}    
\hline 
\noalign{\smallskip}   
\multicolumn{1}{c}{Galaxy} &  
\multicolumn{1}{c}{Position} &   
\multicolumn{1}{c}{$\nabla$Age} &
 \multicolumn{1}{c}{$\nabla\log$(Age)} &
\multicolumn{1}{c}{$\nabla$\ZH} &   
\multicolumn{1}{c}{$\nabla$\aFe} \\   
\multicolumn{1}{c}{} &  
\multicolumn{1}{c}{} &   
\multicolumn{1}{c}{[Gyr $r_{\rm e}^{-1}$]} &
 \multicolumn{1}{c}{[dex decade$^{-1}$]} & 
\multicolumn{1}{c}{[dex decade$^{-1}$]} & 
\multicolumn{1}{c}{[dex decade$^{-1}$]} \\ 
\multicolumn{1}{c}{(1)} &   
\multicolumn{1}{c}{(2)} &   
\multicolumn{1}{c}{(3)} & 
\multicolumn{1}{c}{(4)} & 
\multicolumn{1}{c}{(5)} & 
\multicolumn{1}{c}{(6)} \\ 
\noalign{\smallskip}   
\hline
\noalign{\smallskip}       
NGC~6482      & MJ & $7\pm2$ & $0.23\pm0.09$ & $-0.44\pm0.08$ & $-0.06\pm0.06$\\ 
              & DG & $8\pm3$ & $0.28\pm0.10$ & $-0.34\pm0.06$ & $-0.03\pm0.04$\\  
              & MN & $6\pm1$ & $0.23\pm0.11$ & $-0.35\pm0.06$ & $-0.06\pm0.04$\\  
\multicolumn{2}{l}{Weighted mean} & $6.5\pm0.9$ & $0.25\pm0.06$ &$-0.37\pm0.04$ & $-0.05\pm0.03$\\   
NGC~7556      & MJ & $4\pm4$ & $0.14\pm0.09$ & $-0.29\pm0.11$ & $-0.02\pm0.08$\\  
              & DG & $4\pm1$ & $0.15\pm0.09$ & $-0.28\pm0.06$ & $-0.02\pm0.03$\\  
              & MN & $2\pm1$ & $0.08\pm0.08$ & $-0.23\pm0.07$ & $0.00\pm0.03$\\  
\multicolumn{2}{l}{Weighted mean} & $3.2\pm0.8$ & $0.13\pm0.05$ & $-0.26\pm0.04$ & $-0.01\pm0.02$\\   
\noalign{\smallskip}       
\hline
\end{tabular} 
\end{small}
\end{center}     
\tablefoot{(1) Galaxy name. (2) Slit position: MJ = major axis, MN =
  minor axis, DG = diagonal axis. (3), (4) Age gradient. (4)
  Metallicity gradient. (5) $\alpha/$Fe enhancement gradient.}
\end{table*}    
%%%%%%%%%%%%%%%%%%%%%%%%%%%%%%%%%%%%%%%%%%%%%%%%%%%%%%%%%%%%%%%%%%%%  

To quantify the gradients of age, metallicity, and $\alpha/$Fe
enhancement along the observed axes of NGC~6482 and NGC~7556, we
analyzed the stellar population parameters obtained from the single
stellar population fit of the line-strength indices \Hb, \Mgb, \Fe,
and \MgFe.
Following \citet{Mehlert2003}, we assumed a power-law radial profile
for \Hb, \Mgb, and \Fe\ and calculated their best-fitting values at
$0.1 r_{\rm e}$ and $r_{\rm e}$
(Fig.~\ref{fig:gradients_indices}). From the best-fitting values of
\Mgb\ and \Fe , we derived the values of \MgFe\ at $0.1 r_{\rm e}$ and
$r_{\rm e}$. Then, we converted the values of the line-strength
indices into the stellar population parameters at $0.1 r_{\rm e}$ and
$r_{\rm e}$ using the model predictions by
\citet{Thomas2003}. Finally, we obtained the gradient of the age,
metallicity, and $\alpha$/Fe enhancement as the difference between
their values at $r_{\rm e}$ and $0.1 r_{\rm e}$. This is
\begin{eqnarray*}
\begin{tabular}{lll}
$\nabla \log({\rm Age})$ & = & 
   $[\log({\rm Age}_{0.1r_{\rm e}}) - \log({\rm Age}_{\rm e})]/
   \log{(0.1r_{\rm e}/r_{\rm e})}$ \cr
  & = & $\log({\rm Age}_{\rm e}) - \log({\rm Age}_{0.1r_{\rm e}})$, \cr
  $\nabla [Z/{\rm H}]$ & = &
  $[\log([Z/{\rm H}]_{0.1r_{\rm e}}) - \log([Z/{\rm H}]_{\rm e})]/
  \log{(0.1r_{\rm e}/r_{\rm e})}$ \cr
  & = & $\log([Z/{\rm H}]_{\rm e}) - \log([Z/{\rm H}]_{0.1r_{\rm e}})$, \cr
  $\nabla [\alpha/{\rm Fe}]$ & = &
  $[\log([\alpha/{\rm Fe}]_{0.1r_{\rm e}}) - \log([\alpha/{\rm Fe}]_{\rm e})]/
  \log{(0.1r_{\rm e}/r_{\rm e})}$ \cr
  & = & $\log([\alpha/{\rm Fe}]_{\rm e}) - \log([\alpha/{\rm Fe}]_{0.1r_{\rm e}})$,
\end{tabular}
\end{eqnarray*}
where ${\rm Age}_{0.1r_{\rm e}}$, $[Z/{\rm H}]_{0.1r_{\rm e}}$, and
$[\alpha/{\rm Fe}]_{0.1r_{\rm e}}$ are the parameters calculated at
$0.1r_{\rm e}$ and ${\rm Age}_{\rm e}$, $[Z/{\rm H}]_{\rm e}$, and
$[\alpha/{\rm Fe}]_{\rm e}$ are at $r_{\rm e}$.
These gradients match those commonly derived by assuming power-law
radial profiles for the stellar population parameters, which are
fitted with the linear relations
\begin{eqnarray*}
& \log({\rm Age}(r/r_{\rm e})) = \log({\rm Age}_{\rm e}) + 
 \nabla \log({\rm Age}) \log(r/r_{\rm e}), \\
& [Z/{\rm H}](r/r_{\rm e}) = [Z/{\rm H}]_{\rm e} + 
  \nabla [Z/{\rm H}] \log(r/r_{\rm e}), \\
& [\alpha/{\rm Fe}](r/r_{\rm e}) = [\alpha/{\rm Fe}]_{\rm e} + 
  \nabla [\alpha/{\rm Fe}] \log(r/r_{\rm e}).
\end{eqnarray*}
Our approach makes the measured gradients less sensitive to variations
of the line-strength indices and, as a consequence, of the stellar
population parameters over small radial scales.  We derived the
stellar population gradients separately for each axis by taking into
account the misalignment with respect to the major axis and axial
ratio of the spheroidal component to cover the same galaxy
regions. The uncertainties on the gradients were calculated using
Monte Carlo simulations taking into account the errors on the
available data, as done in \citet{Morelli2012}.  We report the
gradients of age, metallicity, and $\alpha/$Fe enhancement together
with the age difference within $r_{\rm e}$ in
Table~\ref{tab:gradients}.

%%%%%%%%%%%%%%%%%%%%%%%%%%%%%%%%%%%%%%%%%%%%%%%%%%%%%%%%%%%%%%%%%%%% 
% FIGURE: GRADIENTS OF LINE-STRENGTH INDICES
\begin{figure*}[t!]
\centering
\includegraphics[angle=90.0,width=0.46\textwidth]{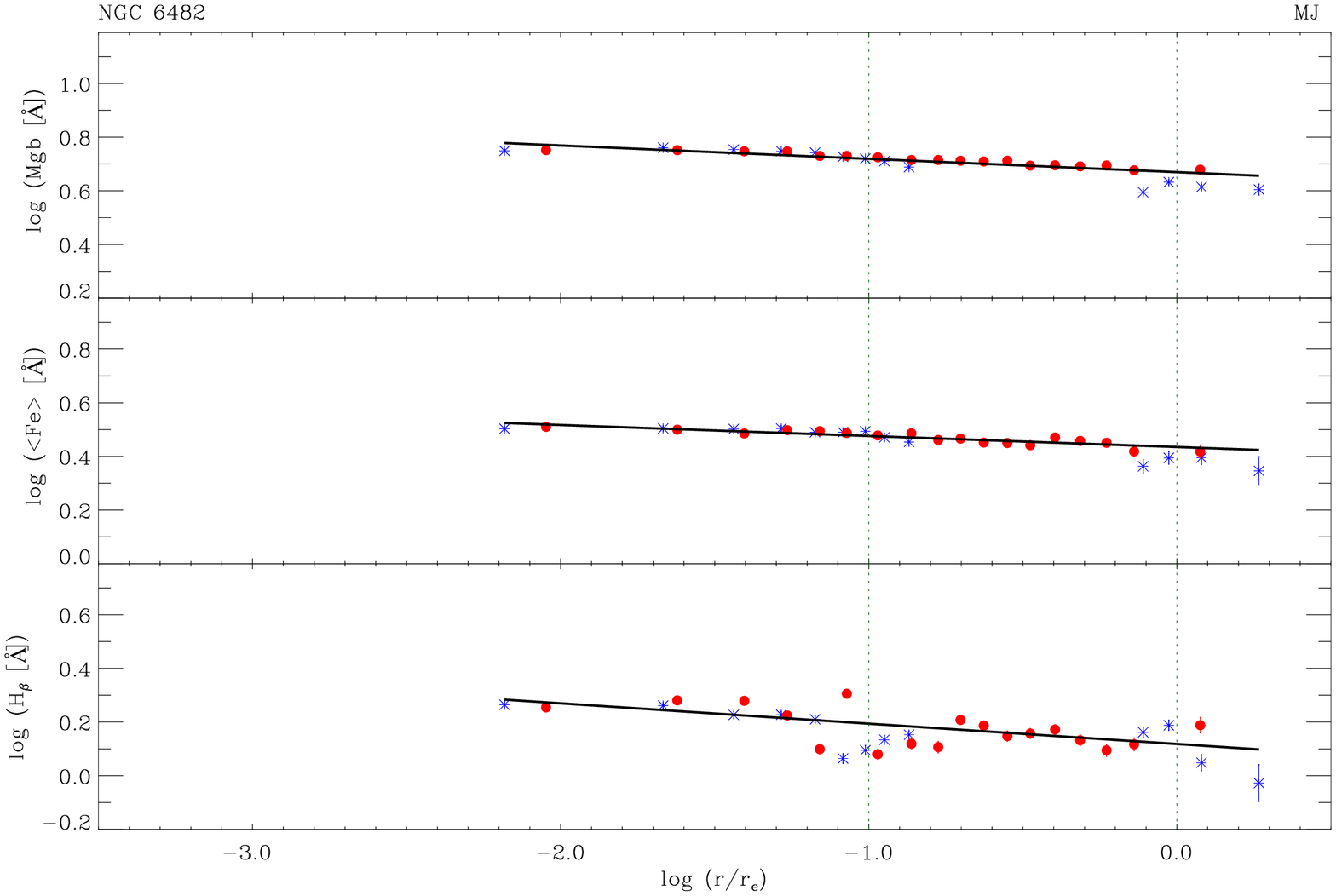}
\includegraphics[angle=90.0,width=0.46\textwidth]{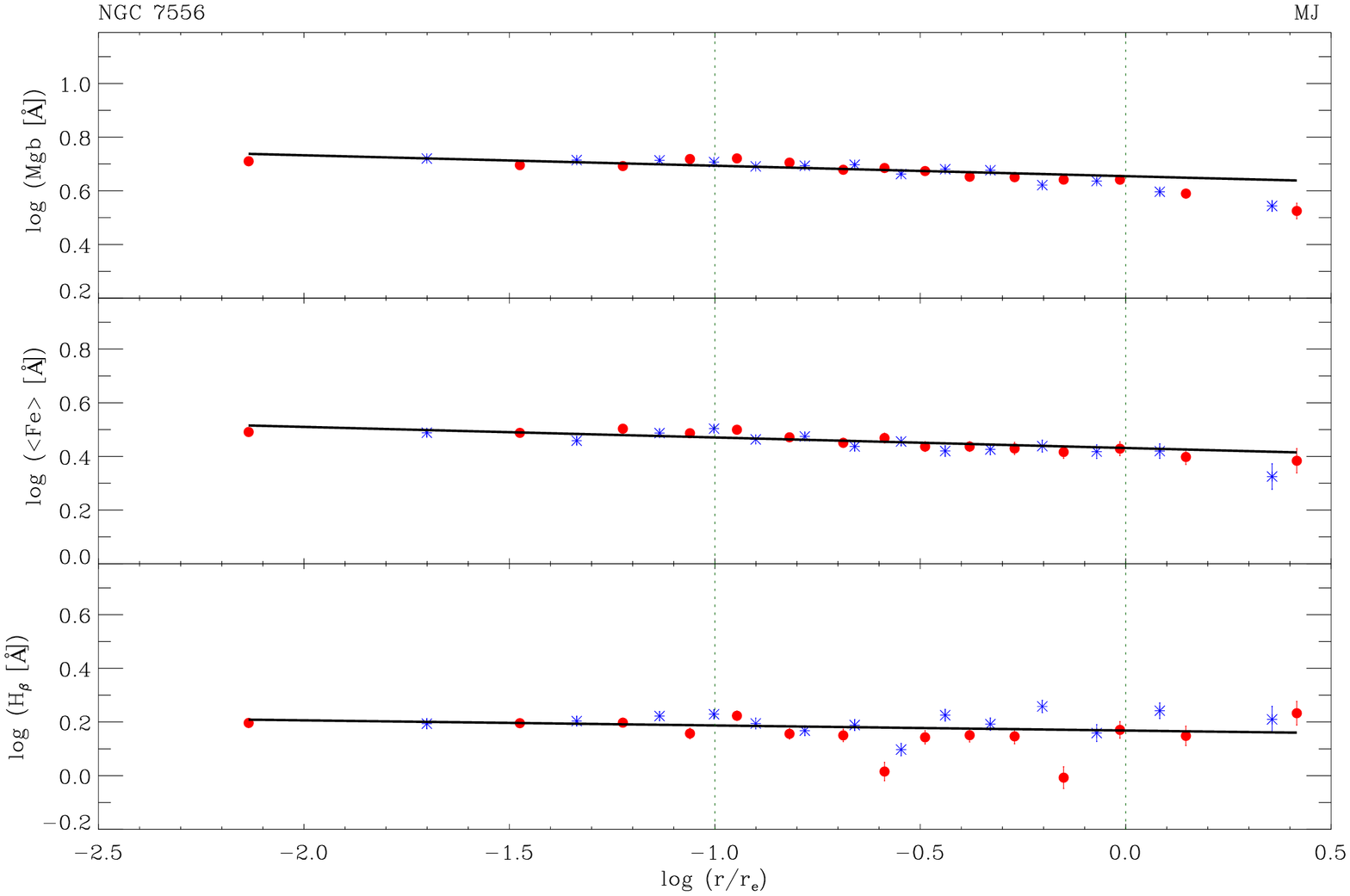}
\includegraphics[angle=90.0,width=0.46\textwidth]{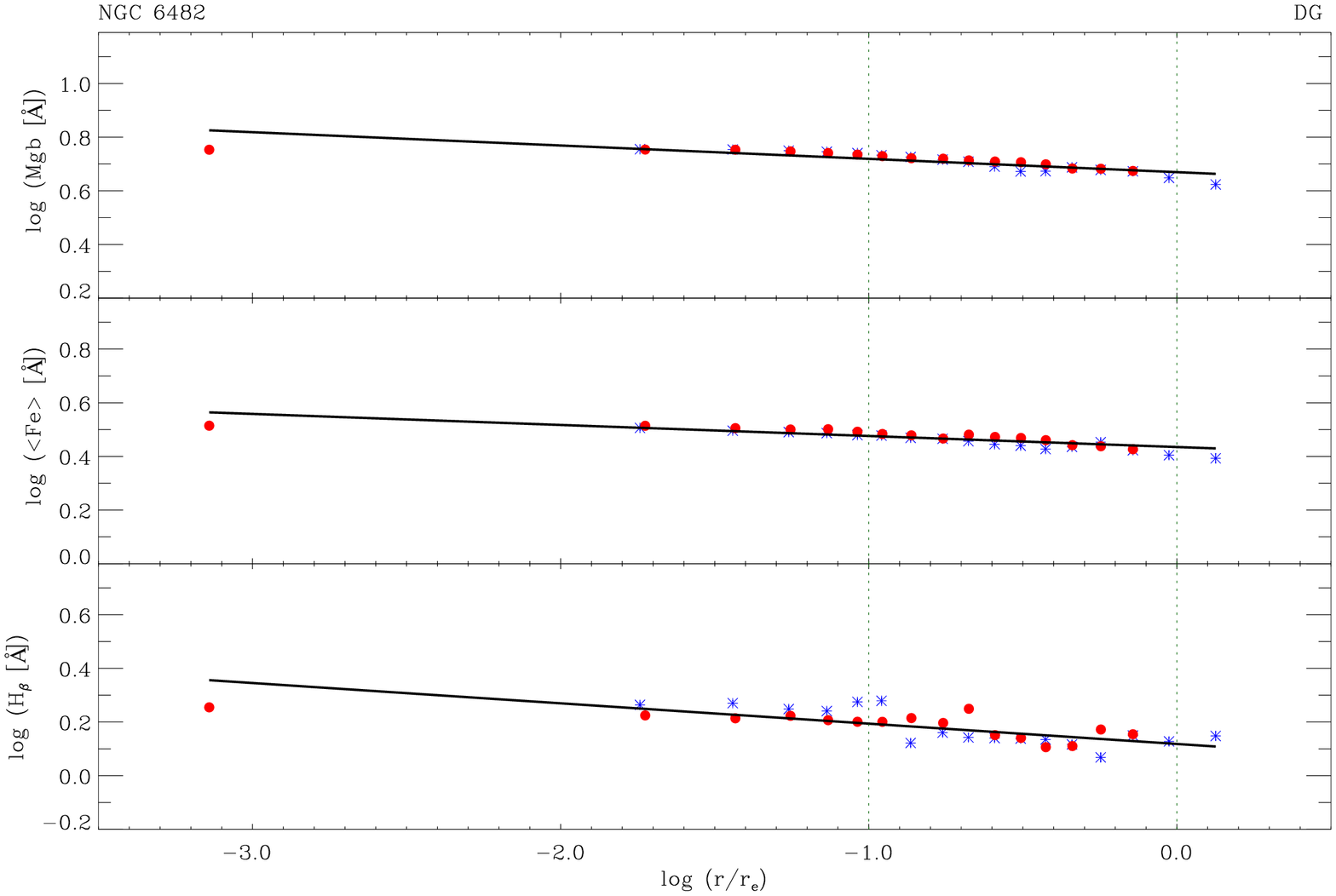}
\includegraphics[angle=90.0,width=0.46\textwidth]{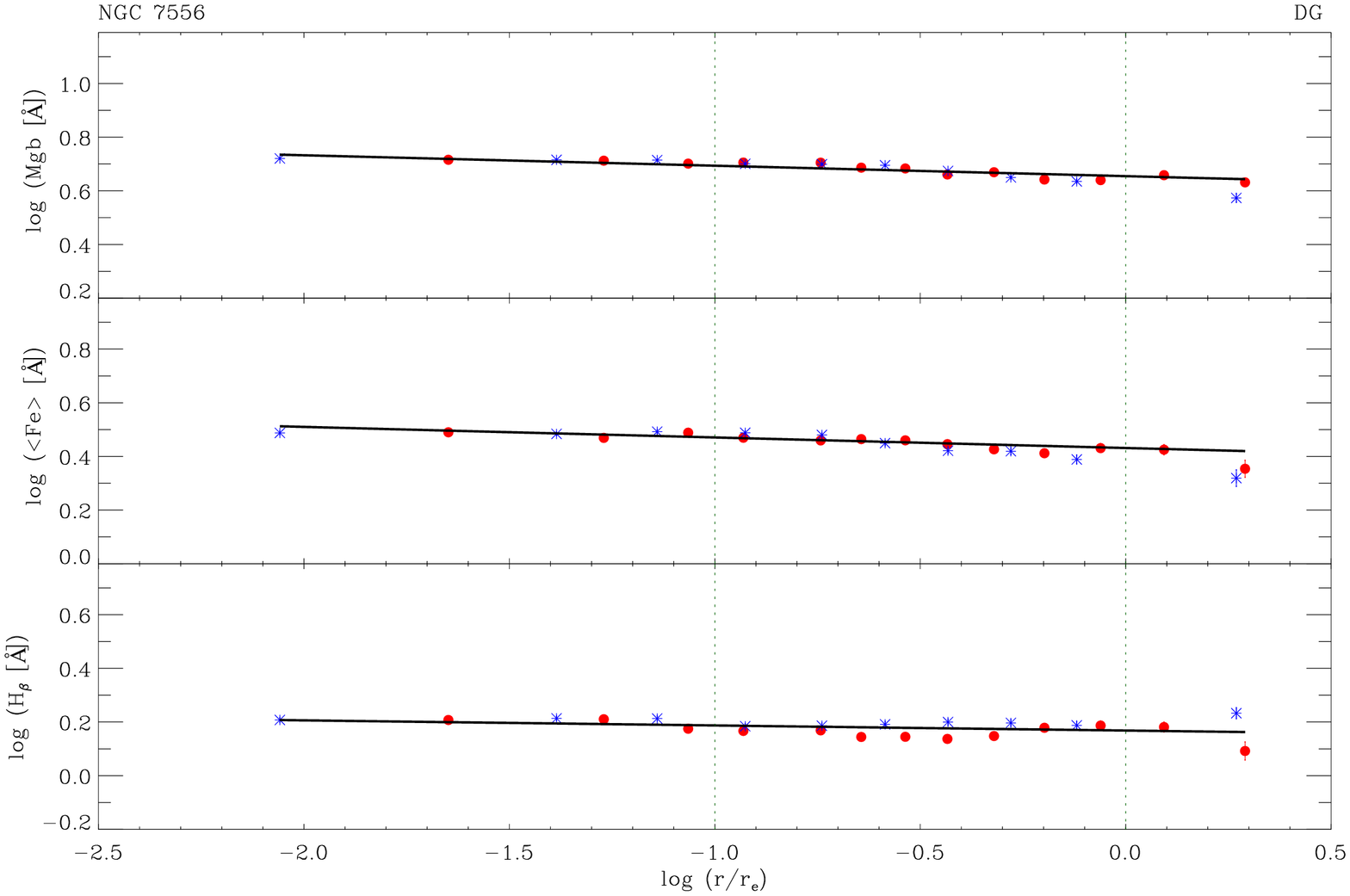}
\includegraphics[angle=90.0,width=0.46\textwidth]{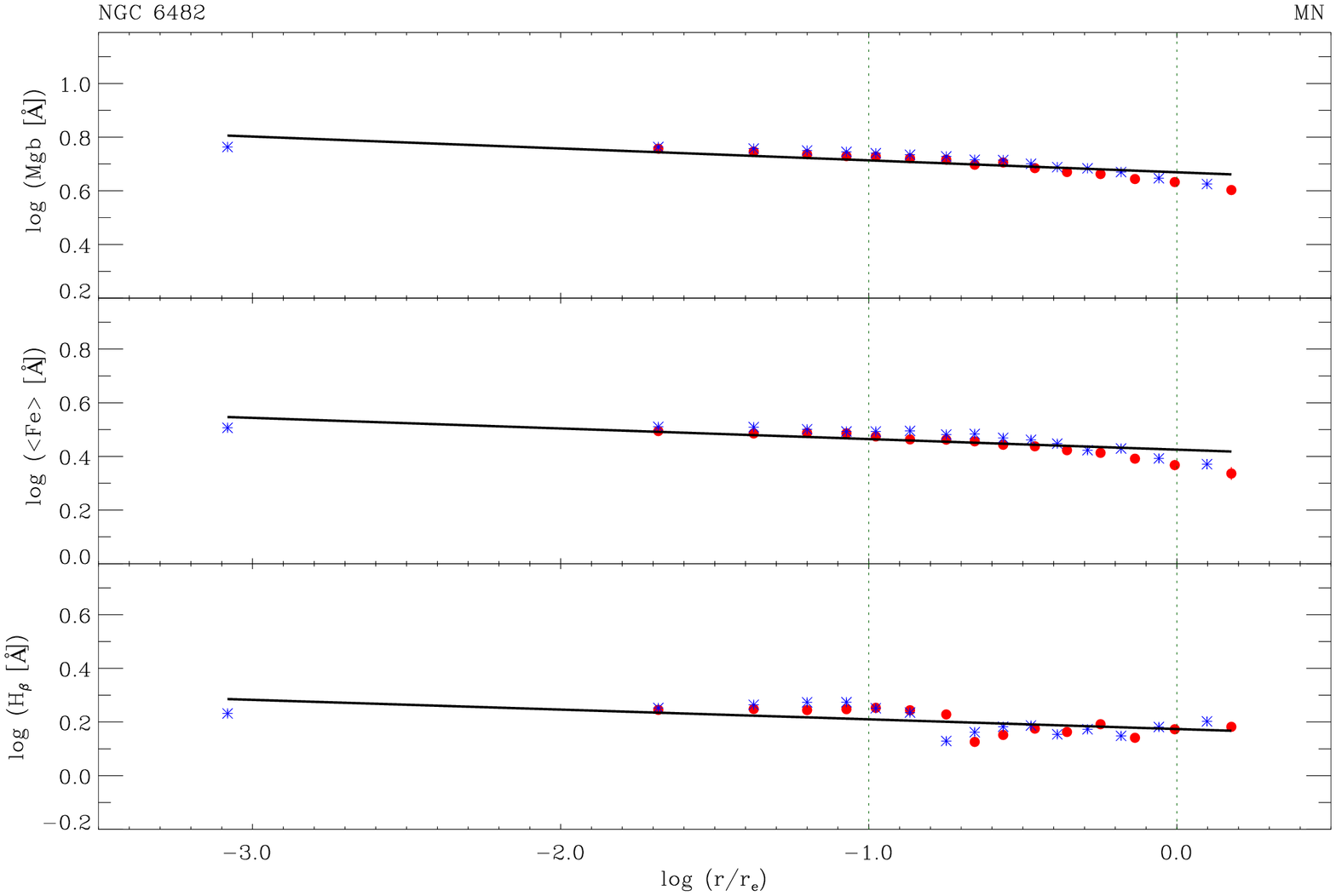}
\includegraphics[angle=90.0,width=0.46\textwidth]{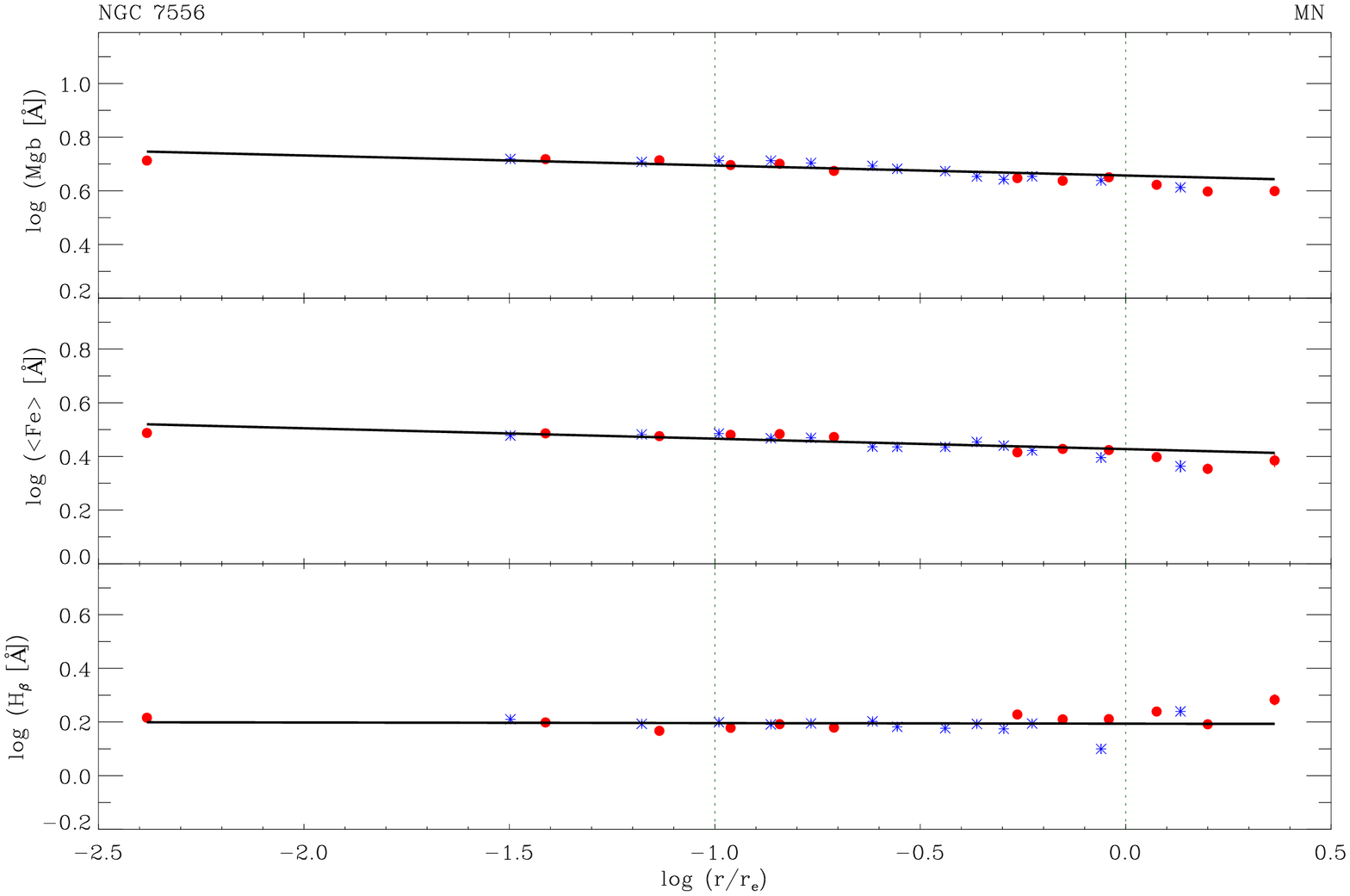}

\caption{Gradients of the line-strength indices \Hb, \Fe, and
  \Mgb\ measured along the major (top panels), diagonal (middle
  panels), and minor axis (bottom panels) of NGC~6482 (left panels) and
  NGC~7556 (right panels). For each axis, the curves are folded around
  the nucleus. Blue asterisks and red circles refer to data measured
  along the approaching and residing sides of the galaxy,
  respectively. The solid line corresponds to the best-fitting power
  law to the available data, while the vertical dotted lines mark the
  radial range between $0.1 r_{\rm e}$ and $r_{\rm e}$ of the
  spheroidal component.  The name of the galaxy and orientation of the
  slit are given for each data set.}
\label{fig:gradients_indices}
\end{figure*}
%%%%%%%%%%%%%%%%%%%%%%%%%%%%%%%%%%%%%%%%%%%%%%%%%%%%%%%%%%%%%%%%%%%% 

The gradients measured on the different axes for the same galaxy are
consistent with each other within errors. This is also true for the
age difference measured within $r_{\rm e}$ along the minor axis of
both galaxies, which at face value seem to be smaller than the
corresponding values measured along the major and diagonal axes.  This
findings confirm the trends we observed for the line-strength indices
and exclude the presence of non-virialized stellar substructures
possibly due to recent merging events.  It was not possible to
completely remove the contribution from the stellar envelope
surrounding the spheroidal component of NGC~7556. However, by limiting
the analysis to $r_{\rm e}$ we ensured a low contamination of the
stellar population of the envelope that dominates the galaxy light for
$r \geq r_{\rm se} > r_{\rm e}$. The envelope contributes from $0.4\%$
to $34\%$ of the total surface brightness moving from the center to
$r_{\rm e}$.

\section{Results and discussion}
\label{sec:discussion}

We used the spatially resolved parameters of the stellar populations
of two FG BCGs, NGC~6482 and NGC~7556, as a benchmark against which
the formation and evolution scenarios of FGs can be tested. We
compared the central values and radial gradients of age and
metallicity of these two galaxies with those of other FG BCGs and
normal ETGs to understand whether they had similar assembly histories.

Figure~\ref{fig:pops_sigma} displays the central metallicity and
central age of NGC~6482 and NGC~7556 against their central velocity
dispersions in comparison with the stellar population properties of
the FG BCGs studied by \citet{Eigenthaler2013} and \citet{Proctor2014}
and the ETGs from \citet{Koleva2011}. To this aim, we converted our
estimates of \ZH\ and \aFe\ into \FeH\ values with $[{\rm Fe/H}] =
[Z/{\rm H}]-0.94[\alpha/{\rm Fe}]$ following \citet{Thomas2003}.
The central velocity dispersions of both NGC~6482 and NGC~7556 are in
the same range of more massive ETGs and FG BCGs. The stellar
populations in the central regions of the two galaxies are younger and
more metal rich than those of the FG BCGs studied so far. The
intermediate age of the central stellar population of NGC~6482 is more
typical for dwarf rather than massive ETGs. The center of NGC~7556 is
slightly younger than BCGs in FGs, but its age falls within the range
observed for the other ETGs and its metallicity is consistent with the
less massive ones.
The central metallicities and central ages plotted in
Fig.~\ref{fig:pops_sigma} are obtained with different methods using
different stellar population models in some
cases. \citet{Eigenthaler2013} and \citet{Proctor2014} explored the
issue of the systematic differences by measuring the stellar
population parameters derived from line-strength indices and full
spectral fitting. They found a fair agreement in most of their FG
BCGs. Unfortunately, \citet{Eigenthaler2013} and \citet{Proctor2014}
do not provide their line-strength indices and a direct comparison
with ours is not possible.

The age and metallicity gradients of FG BCGs are compared to those of
ETGs as a function of their central values of velocity dispersion,
age, and metallicity in Fig.~\ref{fig:gradients_pops}.
The metallicity gradients of both NGC~6482 ($\nabla[{\rm Fe/H}] =
-0.32\pm0.06$ dex decade$^{-1}$) and NGC~7556 ($\nabla[{\rm Fe/H}] =
-0.25\pm0.06$ dex decade$^{-1}$) are within the range observed for
ETGs and FG BCGs. The age gradient of NGC~6482 ($\nabla \log{({\rm
    Age})} = 0.25\pm0.06$ dex decade$^{-1}$) is remarkably larger than
that of FG BCGs, while that of NGC~7556 ($\nabla \log{({\rm Age})} =
0.13\pm0.05$ dex decade$^{-1}$) is marginally consistent with
them. Both values fall within the range of age gradients found for
ETGs.

%%%%%%%%%%%%%%%%%%%%%%%%%%%%%%%%%%%%%%%%%%%%%%%%%%%%%%%%%%%%%%%%%%%% 
% FIGURE: POPULATION PROPERTIES VS SIGMA
\begin{figure}[t!]
\centering
\includegraphics[width=0.49\textwidth]{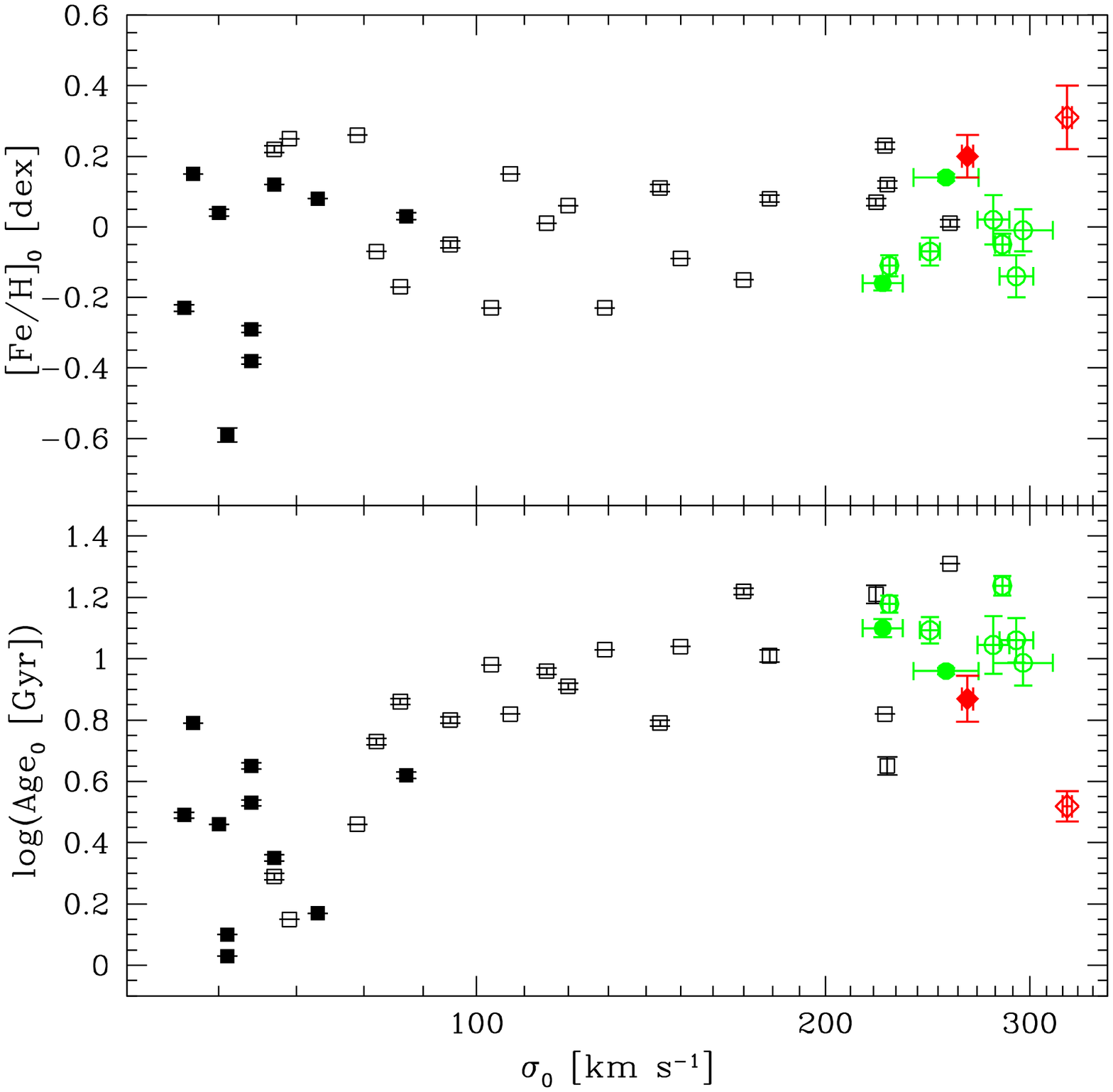}
\caption{Central metallicity (top panel) and central age (bottom
  panel) as a function of the central velocity dispersion for NGC~6482
  (red open diamond) and NGC~7556 (red filled diamond), the FG BCGs
  from \citet[][green open circles]{Eigenthaler2013} and
  \citet[][green filled circles]{Proctor2014}, and the early-type
  normal (open squares) and dwarf (filled squares) galaxies with
  $\sigma>50$ \kms\ from \citet{Koleva2011}.}
\label{fig:pops_sigma}
\end{figure}
%%%%%%%%%%%%%%%%%%%%%%%%%%%%%%%%%%%%%%%%%%%%%%%%%%%%%%%%%%%%%%%%%%%% 

%%%%%%%%%%%%%%%%%%%%%%%%%%%%%%%%%%%%%%%%%%%%%%%%%%%%%%%%%%%%%%%%%%%% 
% FIGURE: POPULATION GRADIENTS VS PROPERTIES
\begin{figure}[t!]
\centering
\includegraphics[width=0.40\textwidth]{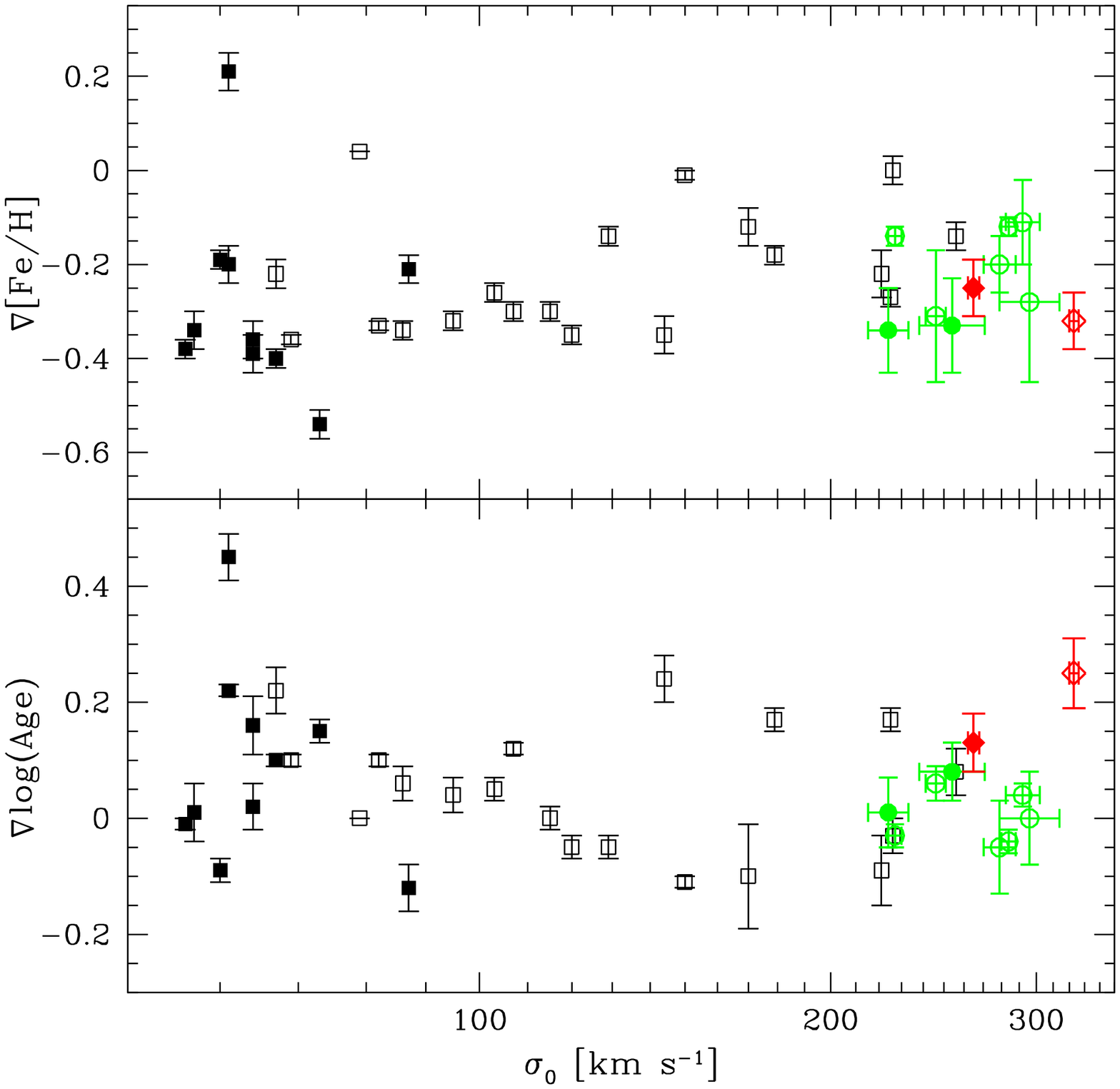}
\includegraphics[width=0.40\textwidth]{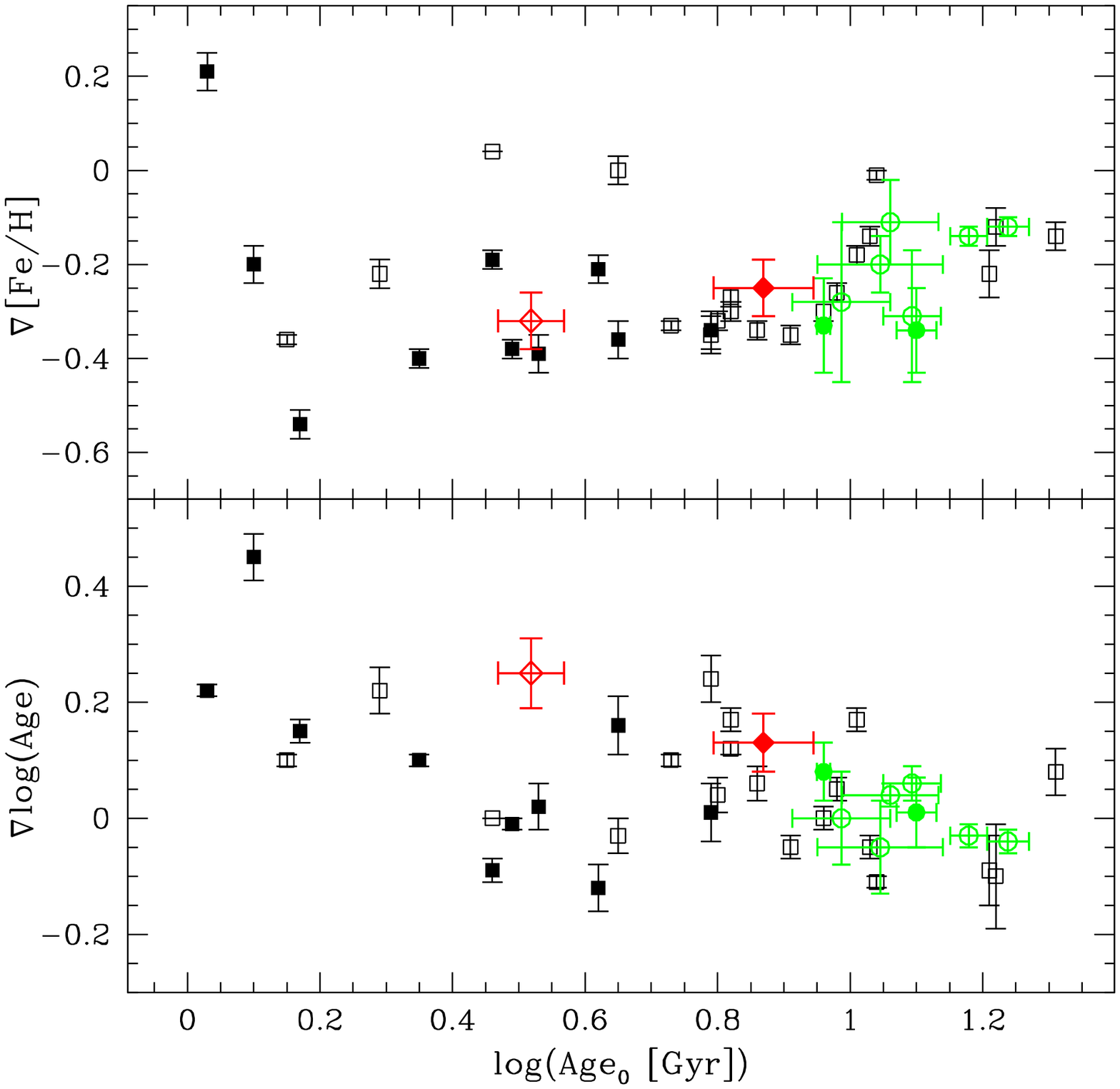}
\includegraphics[width=0.40\textwidth]{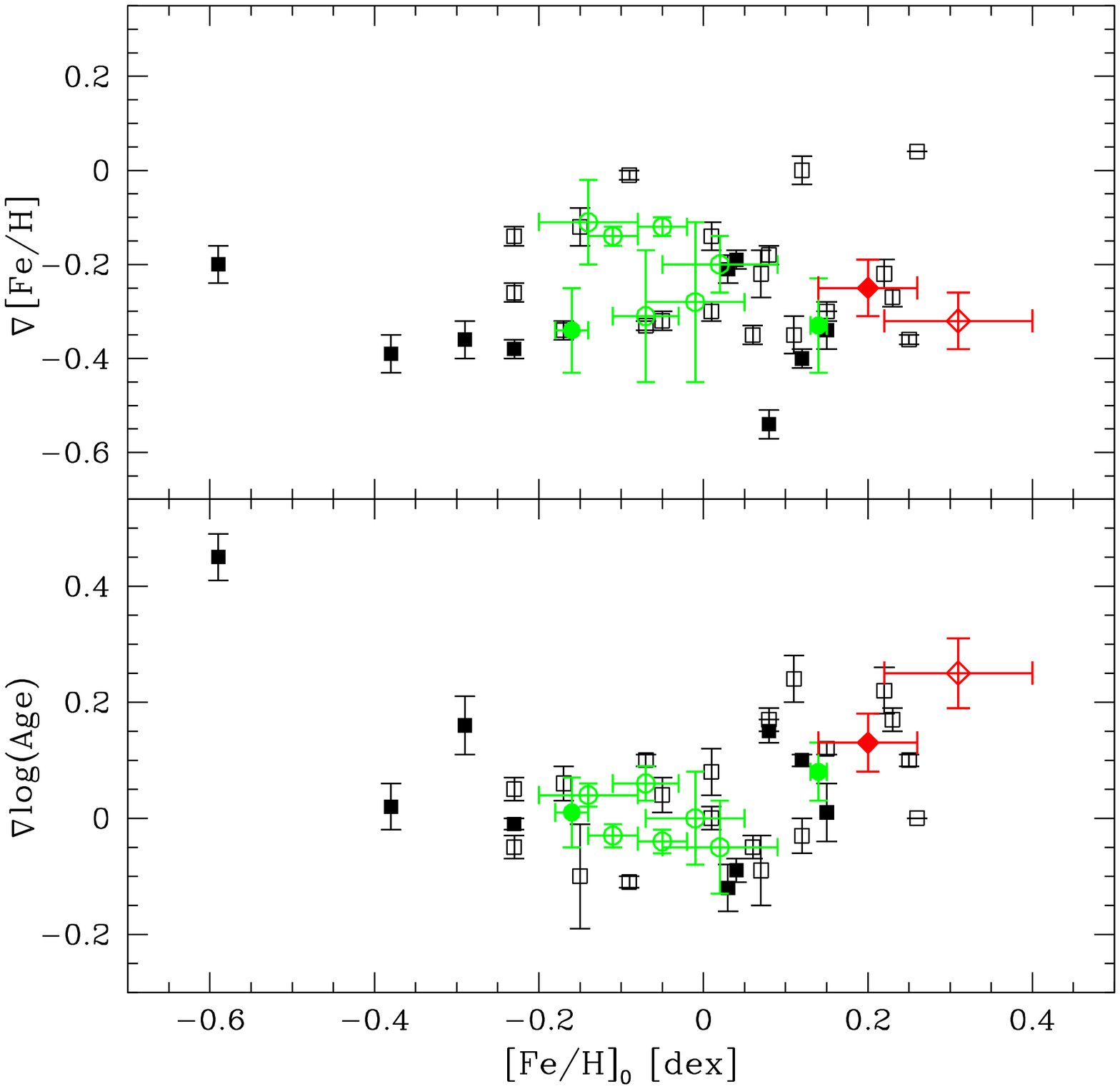}
\caption{Gradients of metallicity and age as a function of the central
  velocity dispersion (top panels) and central age (middle panels) and
  central metallicity (bottom panels) of the galaxies plotted in
  Fig.~\ref{fig:pops_sigma}.}
\label{fig:gradients_pops}
\end{figure}
%%%%%%%%%%%%%%%%%%%%%%%%%%%%%%%%%%%%%%%%%%%%%%%%%%%%%%%%%%%%%%%%%%%% 

This trend is even clearer in Fig.~\ref{fig:gradients}, which shows
that FG BCGs have the same metallicity and age gradients as the bulk
of ETGs when our galaxies are considered together with those of
\citet{Eigenthaler2013} and \citet{Proctor2014}. The FG BCGs are very
massive, in fact the weighted mean of their central velocity
dispersion is $\langle \sigma_0 \rangle = 271\pm1$ \kms . Their
gradients of metallicity and age span over the ranges $-0.1 \lesssim
\nabla[{\rm Fe/H}] \lesssim 0.3$ dex decade$^{-1}$ and $-0.4 \lesssim
\nabla \log{({\rm Age})} \lesssim 0.0$ dex decade$^{-1}$,
respectively. Their weighted means are $\langle \nabla[{\rm Fe/H}]
\rangle = -0.16\pm0.01$ dex decade$^{-1}$ and $\langle \nabla
\log{({\rm Age})} \rangle = 0.01\pm0.01$ dex decade$^{-1}$. The mean
value of the metallicity gradient is driven by NGC~1132 ($\nabla[{\rm
    Fe/H}] = -0.14\pm0.02$ dex decade$^{-1}$) and RX~J0844.9$+$4258
($\nabla[{\rm Fe/H}] = -0.12\pm0.02$ dex decade$^{-1}$). By excluding
these two galaxies, it is $\langle \nabla[{\rm Fe/H}] \rangle =
-0.26\pm0.03$ dex decade$^{-1}$. We also find $\langle \nabla[Z/{\rm
    H}] \rangle = -0.32\pm0.03$ dex decade$^{-1}$ for the FG BCGs for
which $\nabla[Z/{\rm H}]$ is available.

%%%%%%%%%%%%%%%%%%%%%%%%%%%%%%%%%%%%%%%%%%%%%%%%%%%%%%%%%%%%%%%%%%%% 
% FIGURE: AGE GRADIENT VS METALLICITY GRADIENT
\begin{figure}[t!]
\centering
\includegraphics[width=0.49\textwidth]{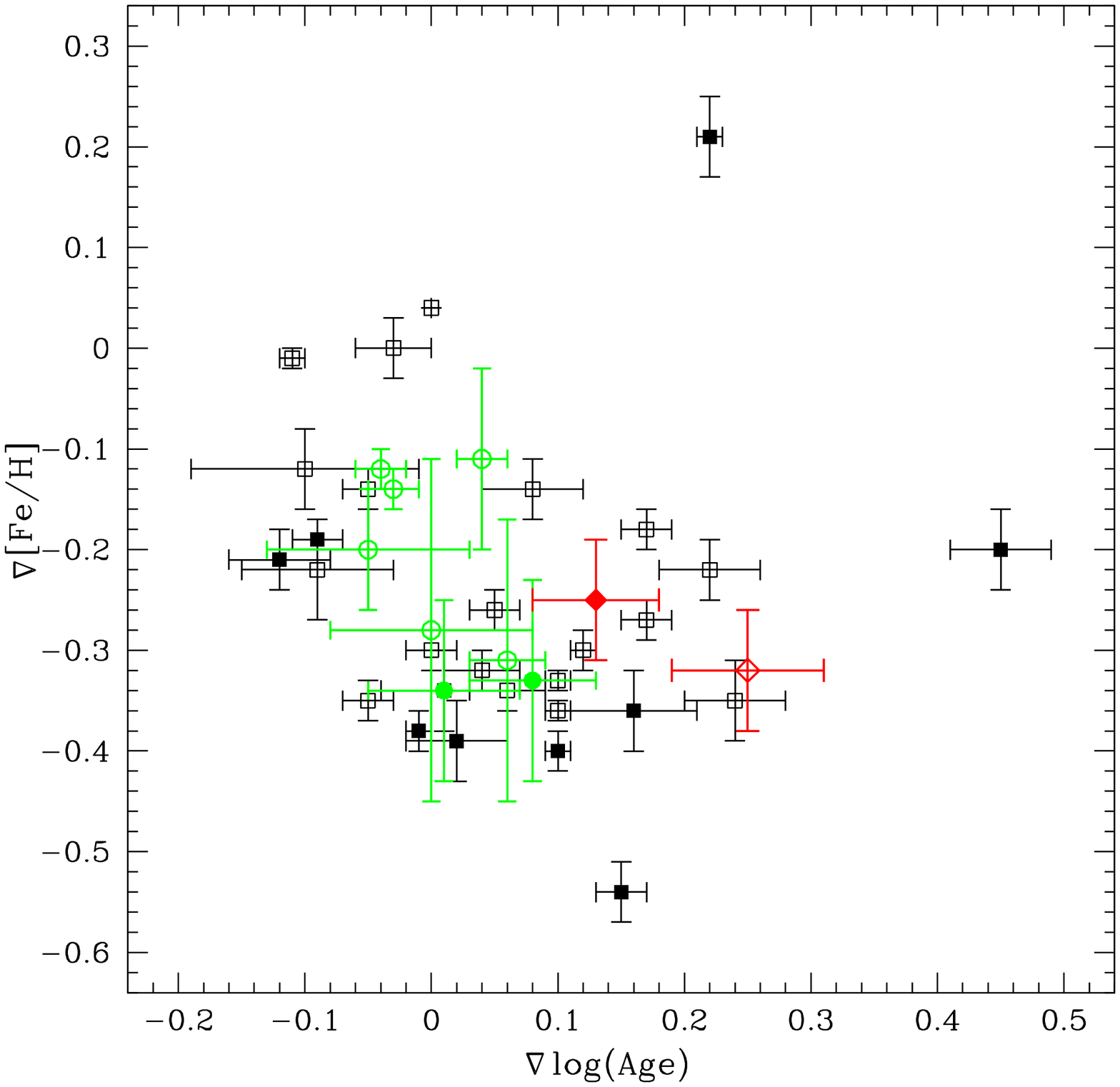}
\caption{Gradient of metallicity as a function of the gradient of age
  of the galaxies plotted in Fig.~\ref{fig:pops_sigma}.}
\label{fig:gradients}
\end{figure}
%%%%%%%%%%%%%%%%%%%%%%%%%%%%%%%%%%%%%%%%%%%%%%%%%%%%%%%%%%%%%%%%%%%% 

Two competing scenarios have been proposed for the formation of FGs in
order to explain the large magnitude gap between the BCGs and their
surrounding galaxies. It has been suggested that FGs are either failed
groups, which formed monolithically from the dissipative collapse of a
pristine gas cloud without bright satellite galaxies and did not
suffer any major merger, or very old systems that assembled
hierarchically and had enough time to exhaust their bright satellite
galaxies through multiple major mergers.

The monolithic formation of ETGs through a pure dissipational collapse
results in steep metallicity gradients ($\nabla [Z/{\rm H}]=-0.5$ dex
decade$^{-1}$, \citealt{Carlberg1984}), since the gas is continuously
enriched by metals ejected from the stars formed as the gas falls
toward the galaxy center. These early findings were confirmed by
\citet{Pipino2008, Pipino2010} by taking into account the gas outflows
triggered by supernova activity and star formation efficiency in
quasi-monolithic collapse models for galaxy formation. Indeed,
\citet{Pipino2010} found metallicity gradients as large as $\nabla
      [Z/{\rm H}] = -0.5$ dex decade$^{-1}$ for a few very massive
      ETGs with $\sigma_0\simeq250$ \kms\ and $M_\ast \simeq
      10^{11.4}$ \msun , which are actually in the same mass range as
      FG BCGs. On the other hand, \citet{Kobayashi2004} investigated a
      further adaptation of monolithic collapse and predicted steep
      metallicity gradients ($\nabla [Z/{\rm H}] =-0.46$ dex
      decade$^{-1}$) also for galaxies assembled from the early
      merging (at $z\gtrsim3$) of many small and gas-rich galaxies
      (with $M_\ast \simeq 10^{9}$ \msun\ and $M_{\rm gas} \geq
      M_\ast$). Present-day galaxies resulting from this
      pseudo-monolithic process can be hardly distinguished from those
      formed from monolithic collapse. The pure monolithic collapse
      and its variants also predict a correlation between the galaxy
      mass and metallicity gradient since a deeper potential retains
      more metals. On the contrary, late major mergers (at
      $z\lesssim3$) are found to dilute the metallicity gradients
      \citep{Kobayashi2004, DiMatteo2009}.

The shallower metallicity gradient measured in FG BCGs with respect to
that predicted from pure monolithic \citep{Carlberg1984}, extreme
quasi-monolithic \citep{Pipino2008}, and pseudo-monolithic
\citep{Kobayashi2004} collapse scenarios suggests that major mergers
played a role in the formation of these very massive galaxies.
Therefore, as alternative routes to form FG BCGs, we considered the
quasi-monolithic collapse in combination with dry major mergers
\citep{DiMatteo2009} and the early assembly of either many small
galaxies (as in the pseudo-monolithic collapse scenario) or a few
large galaxies (with $M_\ast \simeq 10^{10}$ \msun\ and $M_{\rm gas}
\leq 0.7 M_\ast$) followed at $z\lesssim3$ by one or more wet major
mergers \citep{Kobayashi2004}.

Quasi-monolithic collapse predicts steeper metallicity gradients for
more massive galaxies, as well as a larger spread of the metallicity
gradients for more massive galaxies ($-0.5 \lesssim \nabla [Z/{\rm H}]
\lesssim 0$ dex decade$^{-1}$ for $M_\ast \approx 10^{11}$ \msun) with
respect to the less massive ones ($-0.4 \lesssim \nabla [Z/{\rm H}]
\lesssim -0.2$ dex decade$^{-1}$ for $M_\ast \approx 10^{10}$ \msun,
\citealt{Pipino2010}). On the contrary, we found the same scatter of
$\nabla [{\rm Fe/H}]$ for FG BCGs and less massive ETGs and a
remarkably small spread of $\nabla [Z/{\rm H}]$ for the FG BCGs for
which this value is available. If FG BCGs did not suffer any major
merger, then having the same metallicity gradients means that all of
them formed in a very similar way. The large variance of $\nabla
[Z/{\rm H}]$ predicted by \citet{Pipino2010} should be observed for
more massive galaxies also when dry major mergers are
considered. Indeed, they may or may not flatten the metallicity
gradient in the remnant depending on the shape of the metallicity
profiles of the progenitors \citep{DiMatteo2009}. Therefore, observing
similar metallicity gradients in the case of dry major mergers implies
that all FG BCGs had the same merging history. In summary, building FG
BCGs through quasi-monolithic collapse requires a significant fine-tuning
of their evolution history. In addition, no correlation between mass
and metallicity gradient is observed when FG BCGs are compared to
giant and dwarf ETGs, favoring early hierarchical assembly
followed by wet major mergers as the formation scenario for FG BCGs.  It
is worth noting that for massive galaxies the effects of cold gas
accretion in the galactic cores at high redshift mimic a monolithic
collapse \citep[e.g.,][]{Keres2009, Birnboim2011, Dave2011}.

The numerical experiments of galaxy merging by \citet{Kobayashi2004}
show that wet major mergers flatten the metallicity gradient of the
remnant in a way that depends on the gas fraction and stellar mass
ratio of the progenitors. They found a typical value of $\nabla
[Z/{\rm H}] = -0.22$ dex decade$^{-1}$, which is close to what we
measured for FG BCGs. We can conclude that FGs are not failed groups
because their BCGs assembled from the merging of large galaxies.  This
supports the recent results of \citet{Kundert2017}, who have
investigated FGs in the Illustris cosmological hydrodynamical
simulation \citep{Vogelsberger2014}. FGs assembled most of their mass
through mergers before $z\sim0.4$, with massive galaxies accreted
during this period having enough time to merge with the BCG by
today. According to \citet{Kundert2017}, the lack of recent group
accretion has prevented the refilling of the bright satellite
population and allowed for a large magnitude gap to develop within the
past few Gyr.

NGC~6482 is the dominant galaxy of a non-cool core FG
\citep{Khosroshahi2004}, while the galaxy cluster of NGC~7556 hosts a
cool core \citep{Democles2010}. In the past, this feature was
interpreted as due to the different merging history of non-cool and
cool core clusters. According to hydrodynamical simulations
\citep{McCarthy2002, Burns2008}, most cool core systems accreted their
mass slowly over time, with major mergers enhancing the central cool
core. On the other hand, non-cool core systems only experienced major
mergers early in their evolution, which destroyed the embryonic cool
core and produced conditions that prevented its re-formation. In this
picture, non-cool core FGs are the real relics of early structure
formation. However, this scenario can be hardly reconciled with the
presence of relatively young and metal-rich stars in the inner regions
of NGC~6482. They are the signature of a recent merging event which
occurred about 3 Gyr ago, when the induced star formation of
reprocessed gas (i.e., with high metallicity) of external origin
rejuvenated the central stellar population and further increased its
metallicity.
More recent cosmological hydrodynamic simulations show that the
effects of stellar and active galactic nucleus feedback
\citep{Rasia2015} and angular momentum of the merger event
\citep{Hahn2017} are key to explain the diversity of non-cool and cool
core clusters. Simulated cool core clusters have metallicity profiles
that are steeper than those of non-cool core clusters and mergers with
low angular momentum dissolve cool cores, whereas this is not the case
for high angular momentum mergers. This gives further support to the
importance of mergers in building FG BCGs and to the investigation of
their stellar population properties over a much larger radial range.

\section{Conclusions}
\label{sec:conclusions}

We presented long-slit spectroscopic observations along the major,
minor, and diagonal axes of NGC~6482 and NGC~7556, which are the BCGs
of two nearby FGs. The measurements included spatially resolved
stellar kinematics and radial profiles of line-strength indices, which
we converted into age, metallicity, and $\alpha$-element abundance
ratio of the stellar populations using the single
stellar-population models by \citet{Thomas2003}.

Both NGC~6482 and NGC~7556 are very massive ($M_\ast\simeq10^{11.5}$
\msun) and large ($D_{25}\simeq50$ kpc) elliptical galaxies. In spite
of being classified as an early S0 galaxy \citep{RC3}, we found that
NGC~7556 is embedded in a faint and non-rotating spheroidal envelope
rather than in a disk. From the analysis of the spatially resolved
properties of the stellar populations, we argue that NGC~6482 hosts a
central and relatively recent burst of star formation superimposed onto
an old stellar body. NGC~7556 displays a centrally-concentrated
stellar population of intermediate age, which is significantly younger
and more metal rich than the rest of the galaxy. We did not come
across any remarkable difference in the properties of stellar
populations measured along the different axes of NGC~6482 and
NGC~7556. This excludes the presence of non-virialized stellar
substructures in the observed radial range, possibly due to recent
merging events. The age gradients of both galaxies are somewhat larger
than those of the other FG BCGs with stellar population parameters
measured at different radii \citep{Eigenthaler2013, Proctor2014},
whereas their metallicity gradients are similarly negative and
shallow. Moreover, both galaxies have negligible gradients of
$\alpha$-element abundance ratio at supersolar value.
 
The metallicity gradients of all the FG BCGs studied so far are less
steep than those predicted for massive galaxies that formed
monolithically and evolved without experiencing any major merger. This
result holds for different models of galaxy formation based on the
dissipative collapse of pristine gas clouds \citep{Carlberg1984,
  Kobayashi2004, Pipino2010}. In addition, we do not observe a
correlation between metallicity gradient and mass and an increase of
the scatter of the metallicity gradients with mass, when FG BCGs are
compared to giant and dwarf ETGs from \citet{Koleva2011}. FG BCGs have
the same metallicity and age gradients of the bulk of ETGs confirming
early results by \citet{LaBarbera2009}. They found no difference
between the integrated stellar population properties of FG BCGs and
bright field ellipticals and concluded that the FGs might not be a
distinct family of true fossil systems, but rather the final stage of
mass assembly.  All these findings favor a formation scenario of FG
BCGs from wet major mergers following an early hierarchical assembly,
as investigated by \citet{Kobayashi2004} with numerical simulations of
galaxy merging. They reported a typical value of the metallicity
gradient for major merger galaxies, which is close to what we measure
for FG BCGs. Therefore, we conclude that the observed BCGs assembled
through major mergers and their FGs are not failed galaxy groups
that lacked bright satellite galaxies from the beginning. This is in
agreement with the recent findings by \citet{Kundert2017} based on
cosmological hydrodynamic simulations. They showed that the origin of
the magnitude gap and BCGs of FGs depends on the recent accretion
history of the groups and that selecting galaxy groups by their
magnitude gap does not guarantee obtaining either early-formed galaxy
systems or undisturbed central galaxies.

\begin{acknowledgements}

We are indebted to the anonymous referee for a constructive report
that helped us to improve the manuscript.
E.M.C., L.M., and L.C. acknowledge financial support from Padua
University through grants DOR1699945/16, DOR1715817/17, DOR1885254/18,
and BIRD164402/16.  L.C. is also grateful to the Instituto de
Astrof\`\i sica de Canarias for hospitality while this paper was in
progress.  J.A.L.A. and J.M.A. thank the support from the Spanish
Ministerio de Economia y Competitividad (MINECO) through the grants
AYA2013-43188-P and AYA2017-83204-P. S.Z. and M.G. acknowledge
financial support from the University of Trieste through the program
``Finanziamento di Ateneo per progetti di ricerca scientifica
(FRA2015)''. S.Z. is also supported by the grant
PRIN-INAF2014-1.05.01.94.02. E.D. gratefully acknowledges the
hospitality of the Center for Computational Astrophysics at the
Flatiron Institute during the completion of this work.
Part of the data used in this research were acquired through the SDSS
Archive (http://www.sdss.org/). This research also made use of the
HyperLeda Database (http://leda.univ-lyon1.fr/) and NASA/IPAC
Extragalactic Database (NED), which is operated by the Jet Propulsion
Laboratory, California Institute of Technology, under contract with
the National Aeronautics and Space Administration
(http://ned.ipac.caltech.edu/).

\end{acknowledgements}
%%%%%%%%%%%%%%%%%%%%%%%%%%%%%%%%%%%%%%%%%%%%%%%%%%%%%%%%%%%%%%%%%%
%%%%%%%%%%%%%%%%%%%%%%%%%%%%%%%%%%%%%%%%%%%%%%%%%%%%%%%%%%%%%%%%%%

\bibliographystyle{aa} % style aa.bst

\begin{appendix}

\section{Data tables}

%%%%%%%%%%%%%%%%%%%%%%%%%%%%%%%%%%%%%%%%%%%%%%%%%%%%%%%%%%%%%%%%%%%% 
% TABLE: PHOTOMETRY
\renewcommand{\tabcolsep}{3pt}
\renewcommand{\arraystretch}{1.2}
\begin{table*}[h!]  
\caption{Isophotal parameters of NGC~7556. 
\label{tab:ellipse}}
\begin{center}
\begin{small}
\begin{tabular}{rcccc}    
\hline 
\noalign{\smallskip}   
\multicolumn{1}{c}{$a$} &  
\multicolumn{1}{c}{$\mu_r$} &   
\multicolumn{1}{c}{$\epsilon$} & 
\multicolumn{1}{c}{${\it PA}$} \\   
\multicolumn{1}{c}{[arcsec]} &  
\multicolumn{1}{c}{[mag arcsec$^{-2}$]} &   
\multicolumn{1}{c}{} & 
\multicolumn{1}{c}{[$^\circ$]} \\ 
\multicolumn{1}{c}{(1)} &   
\multicolumn{1}{c}{(2)} &   
\multicolumn{1}{c}{(3)} & 
\multicolumn{1}{c}{(4)} \\ 
\noalign{\smallskip}   
\hline
\noalign{\smallskip}       
   0.40 & $17.350\pm0.006$ & $0.158\pm0.042$ & $116.4\pm8.5$\\
   0.79 & $17.513\pm0.006$ & $0.161\pm0.012$ & $111.4\pm2.4$\\
   1.19 & $17.733\pm0.004$ & $0.167\pm0.004$ & $111.5\pm0.8$\\
   1.58 & $17.975\pm0.005$ & $0.178\pm0.003$ & $110.4\pm0.6$\\
   1.98 & $18.208\pm0.005$ & $0.196\pm0.002$ & $109.5\pm0.3$\\
   \dots& \ldots           & \ldots          & \ldots       \\   
\noalign{\smallskip}       
\hline
\end{tabular} 
\end{small}
\end{center}     
\tablefoot{A machine-readable version of the full table is available
   online. A few rows of the table are given as an example. (1)
   Semimajor axis. (2) Ellipse-averaged $r$-band surface
   brightness. (3) Ellipticity defined as $\epsilon = 1 - b/a$ with $b$
   semiminor axis. (4) Major-axis position angle measured north
   through east.}
\end{table*}    
%%%%%%%%%%%%%%%%%%%%%%%%%%%%%%%%%%%%%%%%%%%%%%%%%%%%%%%%%%%%%%%%%%%%  

%%%%%%%%%%%%%%%%%%%%%%%%%%%%%%%%%%%%%%%%%%%%%%%%%%%%%%%%%%%%%%%%%%%% 
% TABLE: KINEMATICS
\renewcommand{\tabcolsep}{3pt}
\begin{table*}[h!]  
\caption{Stellar kinematics of NGC~6482 and NGC~7556.
\label{tab:kinematics}}
\begin{center}
\begin{small}
\begin{tabular}{rcccccr}    
\hline 
\noalign{\smallskip}   
\multicolumn{1}{c}{$r$} &  
\multicolumn{1}{c}{$v$} &   
\multicolumn{1}{c}{$\sigma$} &
\multicolumn{1}{c}{$h_3$} &
\multicolumn{1}{c}{$h_4$} & 
\multicolumn{1}{c}{${\it PA}$} \\   
\multicolumn{1}{c}{[arcsec]} &  
\multicolumn{1}{c}{[\kms]} &   
\multicolumn{1}{c}{[\kms]} & 
\multicolumn{1}{c}{} & 
\multicolumn{1}{c}{} &
\multicolumn{1}{c}{[$^\circ$]} \\ 
\multicolumn{1}{c}{(1)} &   
\multicolumn{1}{c}{(2)} &   
\multicolumn{1}{c}{(3)} &
\multicolumn{1}{c}{(4)} &
\multicolumn{1}{c}{(5)} &
\multicolumn{1}{c}{(6)} \\ 
\noalign{\smallskip}   
\hline
\noalign{\smallskip}       
  \multicolumn{6}{c}{NGC~6482} \\ 
  $-19.3$ & $124.1\pm6.9$ & $340.6\pm7.8$ & $-0.124\pm0.015$ & $ 0.146\pm0.017$ & $+65$\\  
  $-11.5$ & $105.3\pm4.7$ & $328.8\pm5.3$ & $-0.028\pm0.010$ & $ 0.077\pm0.012$ & $+65$\\  
  $ -9.2$ & $105.3\pm4.3$ & $323.8\pm5.2$ & $-0.043\pm0.010$ & $ 0.102\pm0.011$ & $+65$\\ 
  $ -7.5$ & $ 75.0\pm4.5$ & $329.9\pm4.8$ & $-0.049\pm0.010$ & $ 0.061\pm0.011$ & $+65$\\  
  $ -6.1$ & $ 51.0\pm4.4$ & $337.7\pm4.8$ & $-0.014\pm0.009$ & $ 0.062\pm0.010$ & $+65$\\  
  \ldots  & \ldots        & \ldots        & \ldots           & \ldots           & \ldots\\
\noalign{\smallskip}       
\hline
\end{tabular} 
\end{small}
\end{center}     
\tablefoot{A machine-readable version of the full table is available
  online. A few rows of the table are given as an example.  (1)
  Radius. (2) Mean velocity after subtraction of systemic velocity. (3)
  Velocity dispersion. (4) Third-order Gauss–Hermite
  coefficient. (5) Fourth-order Gauss–Hermite coefficient. (6) Slit
  position angle measured north through east.}
\end{table*}    
%%%%%%%%%%%%%%%%%%%%%%%%%%%%%%%%%%%%%%%%%%%%%%%%%%%%%%%%%%%%%%%%%%%%  

%%%%%%%%%%%%%%%%%%%%%%%%%%%%%%%%%%%%%%%%%%%%%%%%%%%%%%%%%%%%%%%%%%%% 
% TABLE: INDICES
\renewcommand{\tabcolsep}{3pt}
\begin{table*}[h!]  
\caption{Line-strength indices of NGC~6482 and NGC~7556.
\label{tab:indices}}
\begin{center}
\begin{small}
\begin{tabular}{rccccccr}    
\hline 
\noalign{\smallskip}   
\multicolumn{1}{c}{$r$} &  
\multicolumn{1}{c}{\Hb} &   
\multicolumn{1}{c}{\Mgd} &
\multicolumn{1}{c}{\Mgb} &
\multicolumn{1}{c}{Fe$_{5270}$} &
 \multicolumn{1}{c}{Fe$_{5335}$} &
\multicolumn{1}{c}{${\it PA}$} \\   
\multicolumn{1}{c}{[arcsec]} &  
\multicolumn{1}{c}{[\AA]} &   
\multicolumn{1}{c}{[mag]} & 
\multicolumn{1}{c}{[\AA]} & 
\multicolumn{1}{c}{[\AA]} &
\multicolumn{1}{c}{[\AA]} &
\multicolumn{1}{c}{[$^\circ$]} \\ 
\multicolumn{1}{c}{(1)} &   
\multicolumn{1}{c}{(2)} &   
\multicolumn{1}{c}{(3)} &
\multicolumn{1}{c}{(4)} &
\multicolumn{1}{c}{(5)} &
\multicolumn{1}{c}{(6)} &
\multicolumn{1}{c}{(7)} \\ 
\noalign{\smallskip}   
\hline
\noalign{\smallskip}       
  \multicolumn{7}{c}{NGC~6482} \\ 
  $-19.3$ & $1.542\pm0.102$ & $0.281\pm0.003$ & $4.772\pm0.130$ & $2.870\pm0.127$ & $2.365\pm0.149$ & $+65$\\
  $-11.5$ & $1.309\pm0.071$ & $0.277\pm0.002$ & $4.748\pm0.095$ & $2.577\pm0.090$ & $2.666\pm0.108$ & $+65$\\
  $ -9.2$ & $1.243\pm0.063$ & $0.284\pm0.002$ & $4.950\pm0.088$ & $2.767\pm0.083$ & $2.879\pm0.099$ & $+65$\\
  $ -7.5$ & $1.355\pm0.064$ & $0.288\pm0.002$ & $4.910\pm0.087$ & $2.810\pm0.081$ & $2.929\pm0.099$ & $+65$\\
  $ -6.1$ & $1.486\pm0.060$ & $0.293\pm0.002$ & $4.957\pm0.090$ & $2.983\pm0.075$ & $2.927\pm0.097$ & $+65$\\
  \ldots  & \ldots          & \ldots          & \ldots          & \ldots          & \ldots          & \ldots\\
\noalign{\smallskip}       
\hline
\end{tabular} 
\end{small}
\end{center}     
\tablefoot{A machine-readable version of the full table is available
  online. A few rows of the table are given as an example.  (1)
  Radius. (2)-(6) Equivalent width of the line-strength indices. (7)
  Slit position angle measured north through east.}
\end{table*}    
%%%%%%%%%%%%%%%%%%%%%%%%%%%%%%%%%%%%%%%%%%%%%%%%%%%%%%%%%%%%%%%%%%%%  

\end{appendix}

\end{document}